\documentclass[reprint,aps,prb,longbibliography,letterpaper,amsmath,final]{revtex4-2}

\usepackage{graphicx}
\usepackage{dcolumn}
\usepackage{bm}
\usepackage[colorlinks = true, linkcolor = blue, urlcolor  = blue, citecolor = blue, anchorcolor = blue]{hyperref}
\usepackage[english]{babel}
\usepackage{makecell}
\usepackage{times}

\begin{document}

\title{Parameter-free analytic continuation for quantum many-body calculations}

\author{Mancheon Han}
\email{einnew90@gmail.com}
\author{Hyoung Joon Choi}
\email{h.j.choi@yonsei.ac.kr}
\affiliation{Department of Physics, Yonsei University, Seoul 03722, Korea}
\date{December 29, 2022}

\begin{abstract}
We develop a reliable parameter-free analytic continuation method for quantum many-body calculations.
Our method is based on a kernel grid, a causal spline, a regularization using the second-derivative
roughness penalty, and the L-curve criterion. We also develop the L-curve averaged deviation to
estimate the precision of our analytic continuation. To deal with statistically obtained data more
efficiently, we further develop a bootstrap-averaged analytic continuation method.
In the test using the exact imaginary-frequency Green's function with added statistical error,
our method produces the spectral function that converges systematically to the exact one as
the statistical error decreases.
As an application, we simulate the two-orbital Hubbard model for various electron numbers with
the dynamical-mean field theory in the imaginary time and obtain the real-frequency self-energy
with our analytic continuation method, clearly identifying a non-Fermi liquid behavior as the
electron number approaches the half filling from the quarter filling. Our analytic continuation can be
used widely and it will facilitate drawing clear conclusions from imaginary-time quantum many-body
calculations.
\end{abstract}

\maketitle

\vspace{-3mm}
\section{Introduction}
\vspace{-3mm}

Numerical simulations of quantum many-body systems in real time often suffer from the
instability caused by oscillatory real-time evolution of $\exp(-iHt)$.
The severe dynamical sign problem in the real-time
quantum Monte Carlo simulation is an example of such instability~\cite{Werner2009,Schiro2010,Cohen2015}.
This instability can be reduced by exploiting the imaginary time by
changing $\exp(-iHt)$ to $\exp(-H\tau)$~\cite{Negele}.
However, it is not straightforward to compare the imaginary-time Green's function
with experimental results,
so the real-frequency Green's function needs to be obtained
from the imaginary-frequency one.
The relation between the imaginary-frequency Green's function $G(i\omega_n)$
and the real-frequency spectral function $A(x)$ is
\begin{equation}
   G(i\omega_n)=\int\frac{A(x)}{i\omega_n-x}dx,\quad  A(x) \geq 0.  \label{eq:ac}
\end{equation}
Using Eq.~(\ref{eq:ac}), one needs to find $A(x)$ from numerically calculated $G(i\omega_n)$.
This procedure is called as the numerical analytic continuation, and it is severely ill-posed~\cite{Silver1990}.

Strong demands for
the numerical analytic continuation led to the development of many methods
despite the ill-posed nature.
Among the various methods, two categories are popular.
One category is estimates of a function $G(z)$ of a complex variable $z$ which interpolates $G(i\omega_n)$.
This category contains the Pad\'e approximant~\cite{Vidberg1977,Deisz1996,Beach2000} 
and the Nevanlinna analytic continuation~\cite{Fei2021}.
The interpolation approach is independent of the real-frequency grid and 
can produce the entire spectral function~\cite{Beach2000,Fei2021}, but it is
very sensitive to numerical precision and the accuracy of input data~\cite{Beach2000}.

The other category is estimates of the spectral function $A(x)$ using
$\chi^2=\sum_{i\omega_n}|G(i\omega_n)-\int\frac{A(x)}{i\omega_n-x}dx|^2$.
One way to use $\chi^2$ is to obtain $A(x)$ by averaging various spectral functions
with the weight of $\exp(-\chi^2)$~\cite{White1991, Sandvik1998, Vafayi2007, Ghanem2020}.
Another way to use $\chi^2$ is to regularize it by adding a regularization parameter $\lambda$
times a functional $R[A]$ so that $A(x)$ is obtained by minimizing $\chi^2+{\lambda}R[A]$.
A representative example of the regularization approach is the 
maximum entropy method~\cite{Silver1990,Gubernatis1991,Jarrell1996,Gunnarsson2010,Kraberger2017,Bergeron2016},
where $R[A]$ is the entropy of the spectral function with respect to the default model $D(x)$.

The regularization approach is stable but its implementations so far require many control parameters.
The maximum entropy method requires the real-frequency grid $\{x_i\}$,
$D(x)$, and $\lambda$.
These parameters can affect the resulting $A(x)$ substantially~\cite{Jarrell1996,Bergeron2016,Wang2009}.
To find optimal parameters, one needs to test several values of $\{x_i\}, D(x),$ and $\lambda$.
While the optimal $\lambda$ can be found with some criteria~\cite{Jarrell1996,Bergeron2016,KaufmannPhD},
methods to determine $\{x_i\}$ and $D(x)$ are not established yet.
In addition, when applied to a metallic system, the maximum entropy method requires
the preblur~\cite{Silver1990,Kraberger2017,Skilling1991,KaufmannPhD}, which
makes it not straightforward to obtain $A(x)$ for metallic and insulating phases on equal footing.

Quantum many-body calculations are often conducted with the imaginary-time quantum Monte Carlo method~\cite{Hirsch1986,Prokofev1998,Rubtsov2005,
Werner2006,Gull2008},
which yields imaginary-frequency data with statistical errors.
To consider statistical errors, it is typical to scale $\chi^2$ with the standard deviation
or the covariance matrix~\cite{Silver1990,Gubernatis1991,Jarrell1996,Bergeron2016},
which can be estimated with resampling methods such as
the jackknife approach~\cite{Kappl2020} or the bootstrap approach~\cite{Efron1994,Chamandy2012}.
Because statistical errors can induce artifacts in the analytically continued spectral function,
the analytic continuation requires careful consideration of statistical errors.

In this work, we develop a reliable parameter-free analytic continuation method.
Our method is based on the regularization approach,
where we remove any arbitrary selection of control parameters as follows.
First, we develop a real-frequency kernel grid which can be used generally
and can support the precise description of corresponding imaginary-frequency data.
Second, we use the second-derivative roughness penalty~\cite{Green1993},
which ensures our method does not need the default model $D(x)$.
Then, the proper regularization parameter $\lambda$ is found by the
L-curve criterion~\cite{Hansen1993,Hansen2001}.
We also develop the L-curve averaged deviation to estimate the precision of our analytic continuation.
In addition, to deal with statistical errors more carefully,
we develop a bootstrap-averaged analytic continuation method.

\vspace{-3mm}
\section{parameter-free analytic continuation}
\vspace{-3mm}

\subsection{Kernel grid}
\vspace{-3mm}

The analytic continuation finds a spectral function $A(x)$ which satisfies Eq.~(\ref{eq:ac})
for given $G(i\omega_n)$.
Numerical implementation of this procedure requires a continuous description of $A(x)$
using a finite number of values.
For this, we used the natural cubic spline~\cite{Boor2,Boor1} interpolation,
where $A(x)$ is represented as
$A(x)=C_{i,0}+C_{i,1}(x-x_i)+C_{i,2}(x-x_i)^2+C_{i,3}(x-x_i)^3$
in the $i$th interval of $x_i{\leq}x{\leq}x_{i+1}$ for $i=1,2,\cdots,n_x-1$, 
with $A''(x_1)=A''(x_{n_x})=0$. Here $n_x$ is the number of grid points
and $A''(x)$ is the second derivative of $A$.
For appropriate grid points, we develop a kernel grid which depends only on
the temperature $k_B T$, the real-frequency cutoff $x_{\textrm{max}}$,
and the number of grid points $n_x$.
The accurate analytic continuation requires the real-frequency grid which 
describes $G(i\omega_n)$ of Eq.~(\ref{eq:ac}) accurately,
so the grid should be dense near $x$ where $A(x)$ contributes
greatly to $G(i\omega_n)$.
Thus, for a single $i\omega_n$, the appropriate grid density should be
proportional to $\left|{\delta G(i\omega_n)}/{\delta A(x)}\right|^2={1}/(\omega_n^2+x^2)$.
Hence, to describe $G(i\omega_n)$ for all $i\omega_n$, we use the grid density $\rho(x)$ such that
\begin{equation}
  \rho(x)\propto \sum_{n=0}^{\infty} \left|\frac{\delta G(i\omega_n)}{\delta A(x)}\right|^2 = \frac{1}{4k_BTx} \tanh\left(\frac{x}{2k_BT}\right), \label{eq:rho}
\end{equation}
where $\omega_n=(2n+1)\pi k_BT$ for the fermionic Green's function.
Then, grid points are determined by the equidistribution principle~\cite{Boor1},
$\int_{x_i}^{x_{i+1}}\rho(x)dx =C/(n_x-1)$ with $x_1=-x_{\textrm{max}}$, $x_{n_x}=x_{\textrm{max}}$,
and $C=\int_{-x_{\textrm{max}}}^{x_{\textrm{max}}}\rho(x)dx$.
Here $x_{\textrm{max}}$ and $n_x$ are determined to be large enough to make the obtained
spectral function converge.

\begin{figure} 
\centering
\centerline{\includegraphics[width=8.5cm]{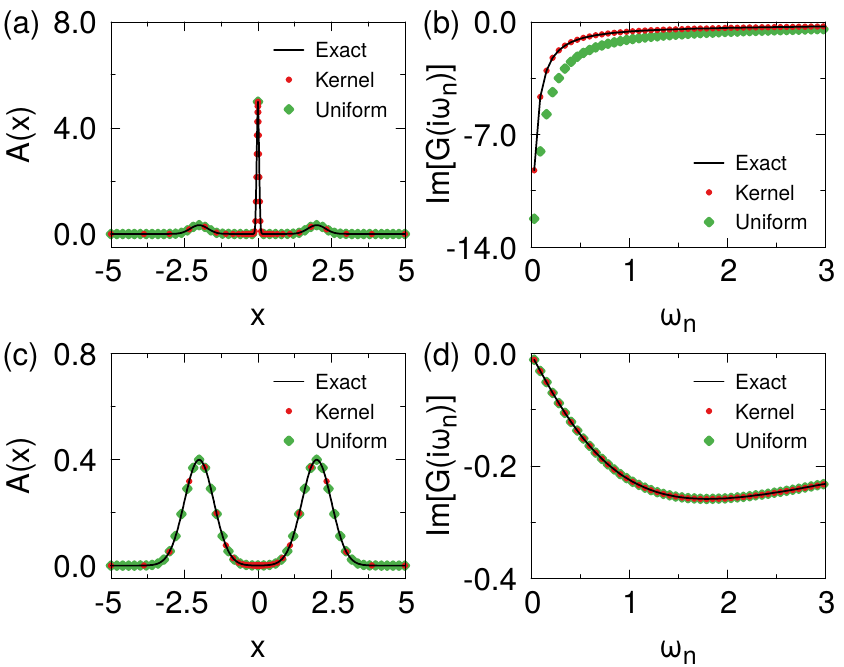}}
\vspace{-3mm}
\caption {Comparison of our kernel grid and a uniform grid in describing the real-frequency spectral function $A(x)$.
(a) and (c) $A(x)$ versus the real frequency $x$.
(b) and (d) The imaginary part of Green's function $G(i\omega_n)$ versus the Matsubara frequency $\omega_n$
corresponding to $A(x)$ shown in (a) and (c), respectively.
Our kernel grid and the uniform grid are generated with $n_x=51$ and $x_{\textrm{max}}=5$.
Temperature is $0.01$. In (a) and (c), values of $A(x)$ at our kernel grid (at the uniform grid) are shown by 
red (green) dots. In (b) and (d), values of the imaginary part of $G(i\omega_n)$ calculated from values of $A(x)$
at our kernel grid (at the uniform grid) are shown by red (green) dots. In (a)\textendash(d),
exact values are shown by black lines.}
\label{fig:1}
\end{figure}

We compare the performance of our kernel grid with that of a uniform grid in Fig.~\ref{fig:1}.
Figures~\ref{fig:1}(a) and \ref{fig:1}(c) show two different spectral functions $A(x)$.
The spectral function $A(x)$ shown in Fig.~\ref{fig:1}(a) has a sharp peak at the Fermi level ($x=0$),
while that shown in Fig.~\ref{fig:1}(c) has no peak at the Fermi level.
Figures~\ref{fig:1}(b) and \ref{fig:1}(d) show $G(i\omega_n)$
corresponding to $A(x)$ shown in Figs.~\ref{fig:1}(a) and \ref{fig:1}(c), respectively.
In the case that $A(x)$ has a sharp peak at the Fermi level ($x=0$), our kernel
grid describes $A(x)$ accurately enough to produce $G(i\omega_n)$ correctly,
while the uniform grid does not [Fig.~\ref{fig:1}(b)].
In the case that $A(x)$ does not have a sharp peak, both our kernel grid
and the uniform grid describe $A(x)$ accurately enough
to produce $G(i\omega_n)$ correctly [Fig.~\ref{fig:1}(d)].

\vspace{-3mm}
\subsection{Causal cubic spline}
\vspace{-3mm}

Finding the spectral function $A(x)$ from $G(i\omega_n)$ by minimizing
$\chi^2[A]=\sum_{i\omega_n}|G(i\omega_n)-\int\frac{A(x)}{i\omega_n-x}dx|^2$
is extremely ill-posed~\cite{Silver1990}.
This ill-posedness can be significantly weakened by imposing the causality condition $A(x)\geq0$~\cite{GhanemPhD}.
Since $A(x_i)\geq0$, $i=1,\cdots,n_x$, satisfies the causality only at the grid points ${x_i}$,
we develop conditions that 
impose the causality for all $x$ as follows.
The cubic spline can be expressed as a linear combination of cubic B-splines
which are non-negative functions~\cite{Lyche2008}.
Thus, $A(x)\geq0$ for all $x$ if expansion coefficients are non-negative~\cite{Lyche2008}: 
\begin{align}
&A(x_1) \geq0,\quad A(x_{n_x}) \geq 0, \nonumber \\
&A(x_i) + \frac{1}{3} (x_{i\pm1}-x_{i}) A'(x_i) \geq 0.
\label{eq:nn}
\end{align}
Here $A'(x)$ is the derivative of $A$.
The cubic spline constrained by Eq.~(\ref{eq:nn}) satisfies $A(x)\geq0$
not only at grid points but also throughout intervals between grid points.
This cubic spline, which we call the \emph{causal} cubic spline, weakens the ill-posedness
of the analytic continuation significantly, but it does not resolve the ill-posedness completely
so minimization of $\chi^2[A]$ with the constraint Eq.~(\ref{eq:nn}) produces $A(x)$ still
having spiky behavior due to overfitting to numerical errors~\cite{GhanemPhD}.
To obtain smooth and physically meaningful $A(x)$, we employ an appropriate roughness penalty
and the L-curve criterion.

\begin{figure} 
\centering
\centerline{\includegraphics[width=8.7cm]{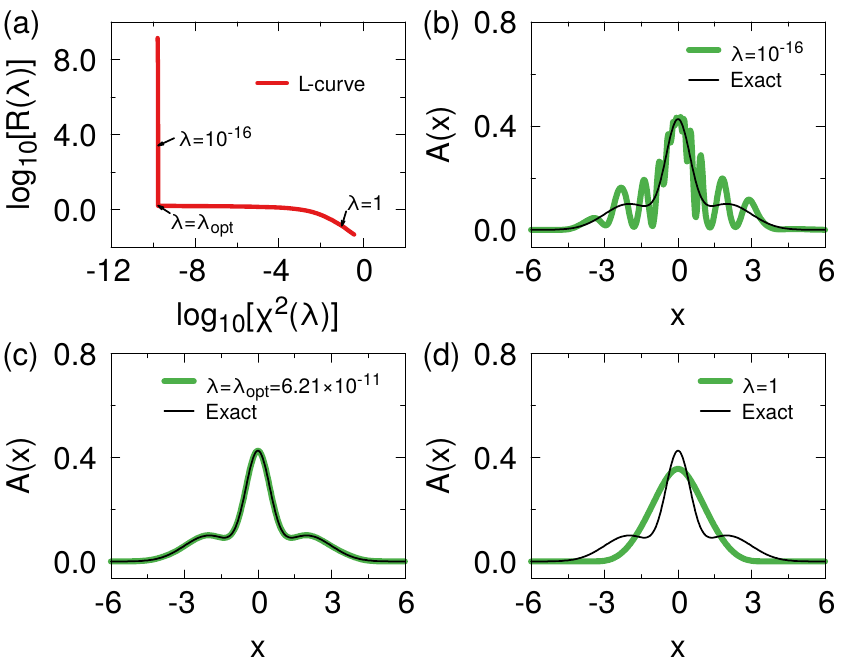}}
\vspace{-3mm}
\caption {The L-curve criterion to find the optimal regularization parameter $\lambda$.
(a) The L-curve. Spectral functions $A(x)$ versus the real frequency $x$, shown by green lines,
computed by minimizing Eq.~(\ref{eq:object}), with (b) $\lambda=10^{-16}$,
(c) $\lambda=\lambda_\textrm{opt}=6.21\times10^{-11}$, and (d) $\lambda=1$.
In (b)\textendash(d), black lines show the exact spectral function for comparison.
The imaginary-frequency Green's function $G(i\omega_n)$ is generated by adding Gaussian
errors with a standard deviation of $10^{-6}$ to the exact one. 
Temperature is $0.01$. We used the first $100$ Matsubara frequencies
and the kernel grid with $x_{\textrm{max}}=10$ and $n_{x}=101$, which are large enough for converged results.}
\label{fig:2}
\end{figure}

\vspace{-3mm}
\subsection{The roughness penalty and the L-curve criterion}
\vspace{-3mm}

To avoid the spiky behavior in $A(x)$ caused by overfitting to numerical errors, 
we use the second-derivative roughness penalty~\cite{Green1993}.
The roughness penalty $R[A]$ is defined as
\begin{equation}
   R[A] = \int_{-x_{\textrm{max}}}^{x_{\textrm{max}}} \left|A''(x)\right|^2 dx.  \label{eq:roughness}
\end{equation}
Then, we obtain $A(x)$ by minimizing a regularized functional
\begin{equation}
   Q_\lambda[A] = \chi^2[A] + \lambda R[A],  \label{eq:object}
\end{equation}
with a regularization parameter $\lambda$.
We use the interior-point method~\cite{Nocedal,mchanCO} to implement this minimization.
Then, the optimal $\lambda$ that balances $\chi^2[A]$ and $R[A]$
is found by the L-curve criterion~\cite{Hansen1993,Hansen2001},
which is a popular approach to determine the regularization parameter in various cases as follows.
For each $\lambda$, one finds $A_\lambda$ that minimizes $Q_\lambda[A]$ in Eq.~(\ref{eq:object}).
Then, let $\chi^2(\lambda)=\chi^2[A_\lambda]$ and $R(\lambda)=R[A_\lambda]$.
The L-curve is the plot of $\operatorname{log}_{10}[R(\lambda)]$ versus $\operatorname{log}_{10}[\chi^2(\lambda)]$.
The L-curve criterion is to choose $\lambda$ that corresponds to the corner of the L-curve
as the optimal value, $\lambda_{\textrm{opt}}$ as illustrated in Fig.~\ref{fig:2}(a).
This procedure can be performed stably and efficiently by
using a recently developed algorithm~\cite{Cultrera2020} which
typically requires minimizations of $Q_\lambda[A]$ at about $20$ different values of $\lambda$.
For a very small $\lambda$, $\chi^2[A]$ dominates $Q_\lambda[A]$ in Eq.~(\ref{eq:object}),
resulting in unphysical peaks in $A_\lambda(x)$, as shown in Fig.~\ref{fig:2}(b).
For a very large $\lambda$, $R[A]$ dominates $Q_\lambda[A]$ in Eq.~(\ref{eq:object}),
resulting in too much broadening in $A_\lambda(x)$, as shown in Fig.~\ref{fig:2}(d).
On the other hand, the spectral function computed with $\lambda_{\textrm{opt}}$ matches excellently
the exact one, as shown in Fig.~\ref{fig:2}(c). 
With this criterion for $\lambda$ 
and with large enough values of $n_x$ and $x_{\textrm{max}}$ 
(see Appendix~\ref{appendix:A} for the convergence test with respect to $n_x$ and $x_{\textrm{max}}$), 
our analytic continuation does not have any arbitrarily chosen parameter that can affect the
real-frequency result significantly, so we call our method a \emph{parameter-free} method.
Our analytic continuation method can be applied to
the self-energy or other Matsubara frequency quantities which
can be represented in a way similar to Eq.~(\ref{eq:ac}).

We can use the L-curve to estimate the precision of the analytic continuation as well.
In the L-curve, $\lambda<\lambda_{\textrm{opt}}$ produces $A_{\lambda}(x)$
which is more fitted to $G(i\omega_n)$, so
the precision of $A_{\lambda_{\textrm{opt}}}(x)$ can be estimated by comparing it with
$A_{\lambda}(x)$.
In this regard, we define the L-curve averaged deviation (LAD),
\begin{align}
    \textrm{LAD}(x) = \frac{\int_{C} [A_{\lambda}(x)-A_{\lambda_\textrm{opt}} (x)]ds}{\int_{C} ds},
\end{align}
where $C$ is the L-curve from $\lambda=0$ to $\lambda=\lambda_{opt}$, and we use it as an error estimator.

\begin{figure}
\centering
\centerline{\includegraphics[width=8.5cm]{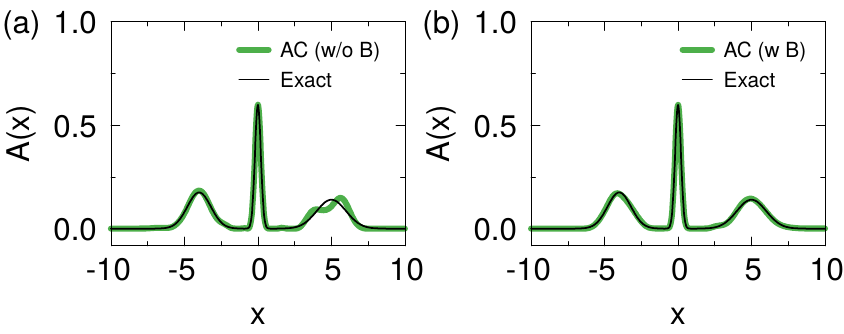}}
\vspace{-3mm}
\caption{Comparison of the spectral functions $A(x)$ from our analytic continuation method without and with the bootstrap average.
(a) $A(x)$ obtained without the bootstrap average.
(b) $A(x)$ obtained with the bootstrap average.
See the text for detailed procedures of our analytic continuation without and with the bootstrap average.
In (a) and (b), green lines are $A(x)$ obtained by our analytic continuation
and black lines are the exact spectral function $A_\textrm{exact}(x)$.}
\label{fig:3}
\end{figure}

\vspace{-3mm}
\subsection{Bootstrap-averaged analytic continuation}
\vspace{-3mm}

The Monte Carlo approach~\cite{Hirsch1986,Prokofev1998,Rubtsov2005,Werner2006,Gull2008}
is often used to calculate the imaginary-frequency Green's function $G(i\omega_n)$.
As a result, the calculated $G(i\omega_n)$ has statistical errors.
The spectral function calculated by the analytic continuation of such data can exhibit artifacts
from statistical errors in $G(i\omega_n)$ [see Fig.~\ref{fig:3}(a) for an example].
Here we devise a bootstrap-averaged analytic continuation method. 
The bootstrap approach~\cite{Efron1994,Chamandy2012} is a widely used resampling method in statistics.
Suppose we have $N$ independent data $G(i\omega_n)_j$, where ${j=1,\cdots,N}$,
and we repeat the bootstrap sampling $N_{\textrm{B}}$ times.
For the $k$th bootstrap sampling, we randomly sample $N$
data from $G(i\omega_n)_j$ with replacement and calculate
their average $g_k^{\textrm{B}}(i\omega_n)=\frac{1}{N}\sum_{j=1}^{N}n_{kj}G(i\omega_n)_j$,
where $n_{kj}$ is the number of repetitions of $G(i\omega_n)_j$ in the $k$th bootstrap sampling.
Then, we obtain the analytically continued spectral function $A[g_k^{\textrm{B}}]$
for $g_k^{\textrm{B}}$.
Finally, the spectral function $A(x)$ is calculated by
$A(x)=\frac{1}{N_\textrm{B}}\sum_{k=1}^{N_\textrm{B}}A[g_k^{\textrm{B}}](x)$,
which converges as $N_\textrm{B}$ increases. In our present work, we used $N_\textrm{B}=256$,
which is large enough to obtain converged results.
Similarly, we can also obtain bootstrap-averaged $\textrm{LAD}(x)$ for the error estimation
of bootstrap-averaged $A(x)$.

\begin{figure} 
\centering
\centerline{\includegraphics[width=8.7cm]{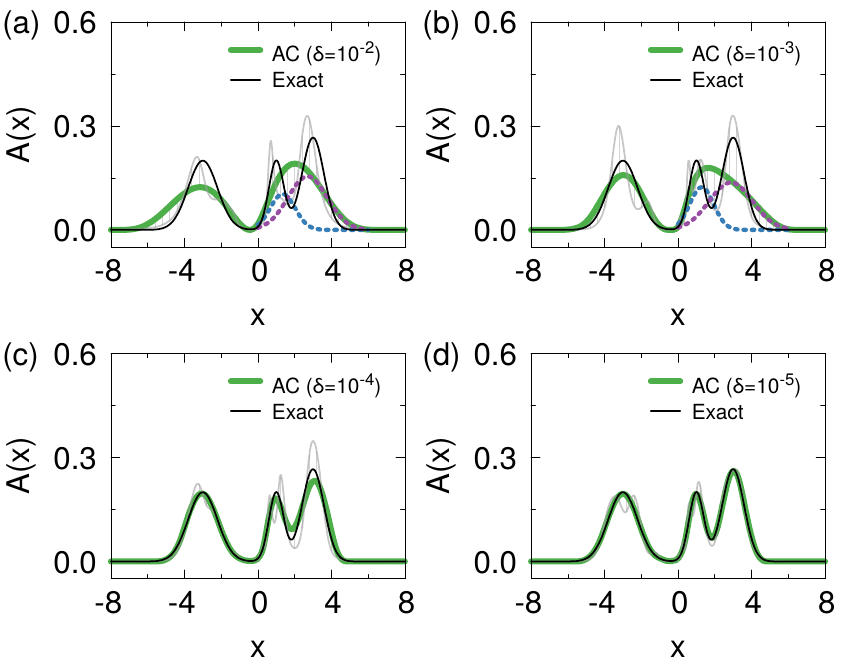}}
\vspace{-3mm}
\caption{The statistical-error dependence of the spectral function $A(x)$, shown by green lines, obtained with
our bootstrap-averaged analytic continuation method.
The standard deviation $\delta$ of the statistical error in imaginary-frequency data is
(a) $10^{-2}$, (b) $10^{-3}$, (c) $10^{-4}$, and (d) $10^{-5}$.
Black lines show the exact spectral function, and gray lines show the deviation of $A(x)$ by $\textrm{LAD}(x)$.
In (a) and (b), Gaussian fits for $A(x)$ are shown in dotted lines.
Temperature is $0.01$. We used the first $100$ Matsubara frequencies
and a kernel grid with $x_{\textrm{max}}=10$ and $n_{x}=101$, which are large enough for converged results.}
\label{fig:4}
\end{figure}

Figure~\ref{fig:3} compares the results of our analytic continuation without and with the bootstrap average.
We consider an exact spectral function $A_\textrm{exact}(x)$ which has a narrow peak at $x=0$ and
a broad peak at $x=-4$ and another broad peak at $x=5$, as shown by the black line in Fig.~\ref{fig:3}.
We obtain the exact Green's function $G_\textrm{exact}(i\omega_n)$ from $A_\textrm{exact}(x)$.
We used a temperature of $0.01$ and the first $100$ Matsubara frequencies. The kernel grid is
generated with $x_\textrm{max}=10$ and $n_x=101$. To perform our analytic continuation without
bootstrap average, we obtain the Green's function $G(i\omega_n)$ by adding a Gaussian error of
the standard deviation of $10^{-6}$ to $G_\textrm{exact}(i\omega_n)$. Then we obtain the
spectral function $A(x)$ by applying our analytic continuation method to $G(i\omega_n)$.
Figure~\ref{fig:3}(a) shows the obtained $A(x)$, which deviates slightly from $A_\textrm{exact}(x)$
at around $x=5$. Next, to perform our bootstrap-averaged analytic continuation,
we obtain $N=100$ independent values of $G(i\omega_n)$ by adding a Gaussian
error of the standard deviation of $10^{-5}$ to $G_\textrm{exact}(i\omega_n)$.
Then, we obtain the spectral function $A(x)$ by applying our bootstrap-averaged analytic continuation
method with $N_\textrm{B}=256$.
Figure~\ref{fig:3}(b) shows the obtained $A(x)$, which agrees excellently with $A_\textrm{exact}(x)$.
The deviation of $A(x)$ from $A_\textrm{exact}(x)$ at around $x=5$ in Fig.~\ref{fig:3}(a) is due to 
overfitting of $A(x)$ to $G(i\omega_n)$ with statistical errors,
and it is avoided by the bootstrap average, as shown in Fig.~\ref{fig:3}(b).
So the bootstrap average is useful for avoiding the overfitting to data with statistical errors.
This bootstrap average can be used with any analytic continuation method.

\vspace{-3mm}
\section{Benchmarks and applications}
\vspace{-3mm}

\subsection{Tests with exact results}
\vspace{-3mm}

Figure~\ref{fig:4} demonstrates the statistical-error dependence of
the spectral function $A(x)$ and $\textrm{LAD}(x)$ obtained with our method.
We consider an exact $A(x)$ consisting of three peaks, from which we obtain the exact $G(i\omega_n)$.
Then, we add statistical errors to the exact $G(i\omega_n)$ and apply our method
to obtain $A(x)$ and $\textrm{LAD}(x)$. 
Here the standard deviation $\delta$ of the statistical errors
is independent of $\omega_n$. 
(See Appendix~\ref{appendix:B} for 
$\delta$ varying with $\omega_n$.)
As shown in Fig.~\ref{fig:4}(a),
when statistical errors in $G(i\omega_n)$ are large,
the obtained $A(x)$ is broader than the exact one, and $\textrm{LAD}(x)$ is large.
As the statistical errors are reduced, $A(x)$
converges to the exact one, and $\textrm{LAD}(x)$ diminishes [Figs.~\ref{fig:4}(b)\textendash(d)].
These results show that our method behaves well with respect to the statistical errors,
reproducing the exact spectral function if the statistical errors are small enough.

\begin{table} 
\caption{Analysis of peak centers and peak widths in the spectral functions shown by green lines in Fig.~\ref{fig:4}.
\vspace{-5mm}
}
\setlength{\tabcolsep}{3mm}
\renewcommand{\arraystretch}{1.2}
\begin{center}
\begin{tabular}{c c c c}
\hline\hline
  & Left peak & Middle peak & Right peak \\
$A(x)$ & Center Width & Center Width & Center Width \\
\hline
Fig.~\ref{fig:4}(a) & $ -3.24\quad 1.28\phantom{0} $ & $ 1.34\quad 0.670 $ & $ 2.67\quad 1.04\phantom{0} $ \\
Fig.~\ref{fig:4}(b) & $ -3.01\quad 0.981 $ & $ 1.30\quad 0.657 $ & $ 2.88\quad 1.18\phantom{0} $ \\
Fig.~\ref{fig:4}(c) & $ -3.01\quad 0.813 $ & $ 1.02\quad 0.466 $ & $ 3.05\quad 0.687 $ \\
Fig.~\ref{fig:4}(d) & $ -3.01\quad 0.813 $ & $ 1.00\quad 0.414 $ & $ 3.02\quad 0.609 $ \\
Exact & $ -3.00\quad 0.800 $ & $ 1.00\quad 0.400 $ & $ 3.00\quad 0.600 $ \\
\hline\hline
\end{tabular}
\vspace{-5mm}
\end{center}
\label{tab:1}
\end{table}

\begin{figure}[b] 
\centering
\centerline{\includegraphics[width=8.5cm]{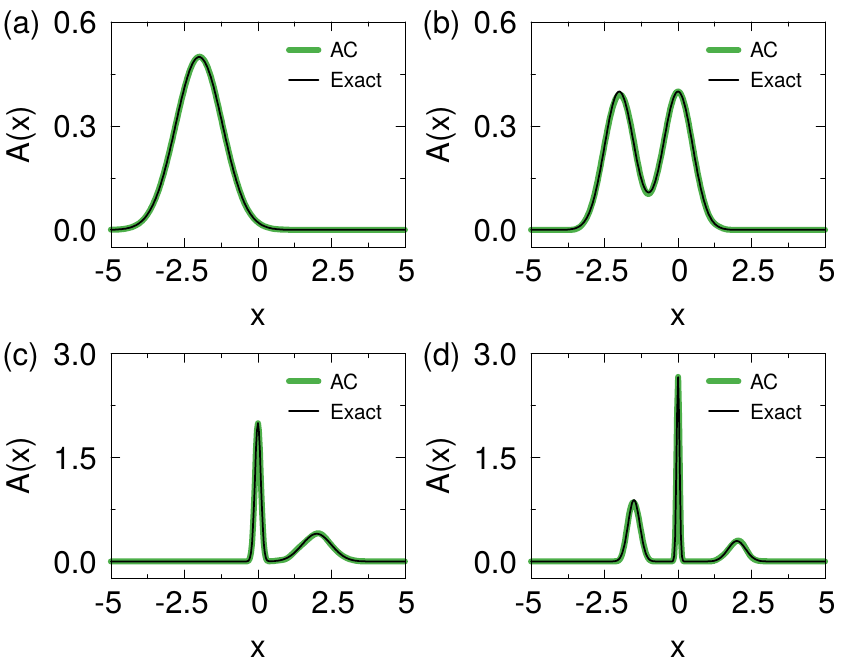}}
\vspace{-3mm}
\caption{Tests of our bootstrap-averaged analytic continuation method for various spectral functions
$A(x)$ consisting of (a) a broad peak at $x<0$, (b)
two broad peaks at $x<0$ and $x=0$, (c)
a narrow peak at $x=0$ and a broad peak at $x>0$, and (d)
a narrow peak at $x=0$ and two broad peaks at $x<0$ and $x>0$.
In (a)\textendash(d), green lines show $A(x)$ obtained with our bootstrap-averaged
analytic continuation, while black lines show exact spectral functions $A_\textrm{exact}(x)$.
We used a temperature of $0.01$, the first $100$ Matsubara frequencies,
and a kernel grid with $x_{\textrm{max}}=10$ and $n_{x}=101$.}
\label{fig:5}
\end{figure}

For a more detailed analysis, we fit the obtained $A(x)$ in Fig.~\ref{fig:4} with three Gaussian functions.
Table~\ref{tab:1} shows the obtained peak centers and peak widths.
These results show explicitly that peak centers and peak widths converge to the corresponding
exact values as statistical errors in $G(i\omega_n)$ decrease and show that
peak centers converge faster than peak widths.

In addition, we tested our analytic continuation method with various spectral functions (Fig.~\ref{fig:5}).
Here we performed the bootstrap average with $N_\textrm{B}=256$, using $N=100$ independent values of
$G(i\omega_n)$. Each independent value of $G(i\omega_n)$ was obtained by adding a Gaussian
error of the standard deviation of $10^{-5}$ to the exact Green's function $G_\textrm{exact}(i\omega_n)$
obtained from the exact spectral function $A_\textrm{exact}(x)$.
Figure~\ref{fig:5} confirms analytically continued spectral functions $A(x)$ agree excellently with
corresponding $A_\textrm{exact}(x)$. 
We also compared our method with the maximum entropy method (Appendix~\ref{appendix:C}).

\vspace{-3mm}
\subsection{Dynamical mean-field theory simulation of the two-orbital Hubbard model}\label{sec:2o_DMFT}
\vspace{-3mm}

As an application of our method, we consider the two-orbital Hubbard model~\cite{Hubbard1963}
described by the Hamiltonian
\begin{align}
H  =&-\sum_{\langle ij\rangle,ab\sigma} t_{ij}^{ab} d^{\dagger}_{ia\sigma} d^{ }_{jb\sigma} -
      \sum_{i\sigma} \mu n_{i\sigma} +  \sum_{ia} U n_{ia\uparrow} n_{ia\downarrow} \nonumber \\
    &+  \sum_{i,a<b,\sigma} \{(U-2J) n_{ia\sigma} n_{ib\bar{\sigma}} + (U-3J) n_{ia\sigma} n_{ib\sigma}\} \nonumber \\
    &-  \sum_{i,a\neq b} J (d^{\dagger}_{ia\downarrow} d^{\dagger}_{ib\uparrow}d^{ }_{ib\downarrow} d^{ }_{ia\uparrow}+
                  d^{\dagger}_{ib\uparrow} d^{\dagger}_{ib\downarrow}d^{ }_{ia\uparrow} d^{ }_{ia\downarrow}) \label{eq:2ohh}
\end{align}
in the infinite-dimensional Bethe lattice with a semicircular noninteracting density of states.
Here $d^{ }_{ia\sigma}$($d^{\dagger}_{ia\sigma}$) is the annihilation (creation) operator of
an electron of spin $\sigma$ in the $a$th ($a=1,2$) orbital at the $i$th site,
$n_{ia\sigma} = d^{\dagger}_{ia\sigma} d^{ }_{ia\sigma}$, $t^{ab}_{ij}$ is the nearest-neighbor
hopping energy, $\mu$ is the chemical potential, $U$ is the local Coulomb interaction,
and $J$ is the Hund's coupling. With the Pad\'e approximant, the imaginary part of the self-energy in this model shows
a peak at around the Fermi level in the non-Fermi-liquid phase~\cite{Hafermann2012}.
Our analytic continuation method makes it possible to analyze
the existence and evolution of peaks in detail, as shown below.

We simulate the two-orbital Hubbard model of Eq.~(\ref{eq:2ohh})
with the dynamical mean field theory (DMFT)~\cite{Metzner1989,Muller1989:a,Muller1989:b,Georges1992,Georges1996}.
We implemented the hybridization-expansion continuous-time quantum Monte Carlo
method~\cite{Werner2006} and used it as an impurity solver.
Both orbitals have the same bandwidth of $4$ in our energy units.
We simulated the case with $U=8, J=U/6,$ and temperature $T=0.02$
and considered the paramagnetic phase.
We applied our analytic continuation method to obtain the real-frequency self-energy $\Sigma^{R}(x)$
from the first $100$ imaginary-frequency self-energies $\Sigma(i\omega_n)$.
We used $n_x=101$ and $x_{\textrm{max}}=30$ to form the kernel grid,
which are large enough for converged results.
For the bootstrap average, we used $1.28\times10^4$ independent sets of $\Sigma(i\omega_n)$
obtained with $2\times10^6$ Monte Carlo steps.
After obtaining $\Sigma^{R}(x)$, we computed the spectral function $A(x)=-\frac{1}{\pi}\operatorname{Im}[G(x+i\eta)]$ by using
$G(x+i\eta)=\int_{-\infty}^{\infty}{D(\epsilon)}/(x+i\eta+\mu-\epsilon-\Sigma^{R}(x))d\epsilon$.
Here $D(\epsilon)=\sqrt{4-x^2}/(2\pi)$.

\begin{figure} 
\centering
\centerline{\includegraphics[width=8.5cm]{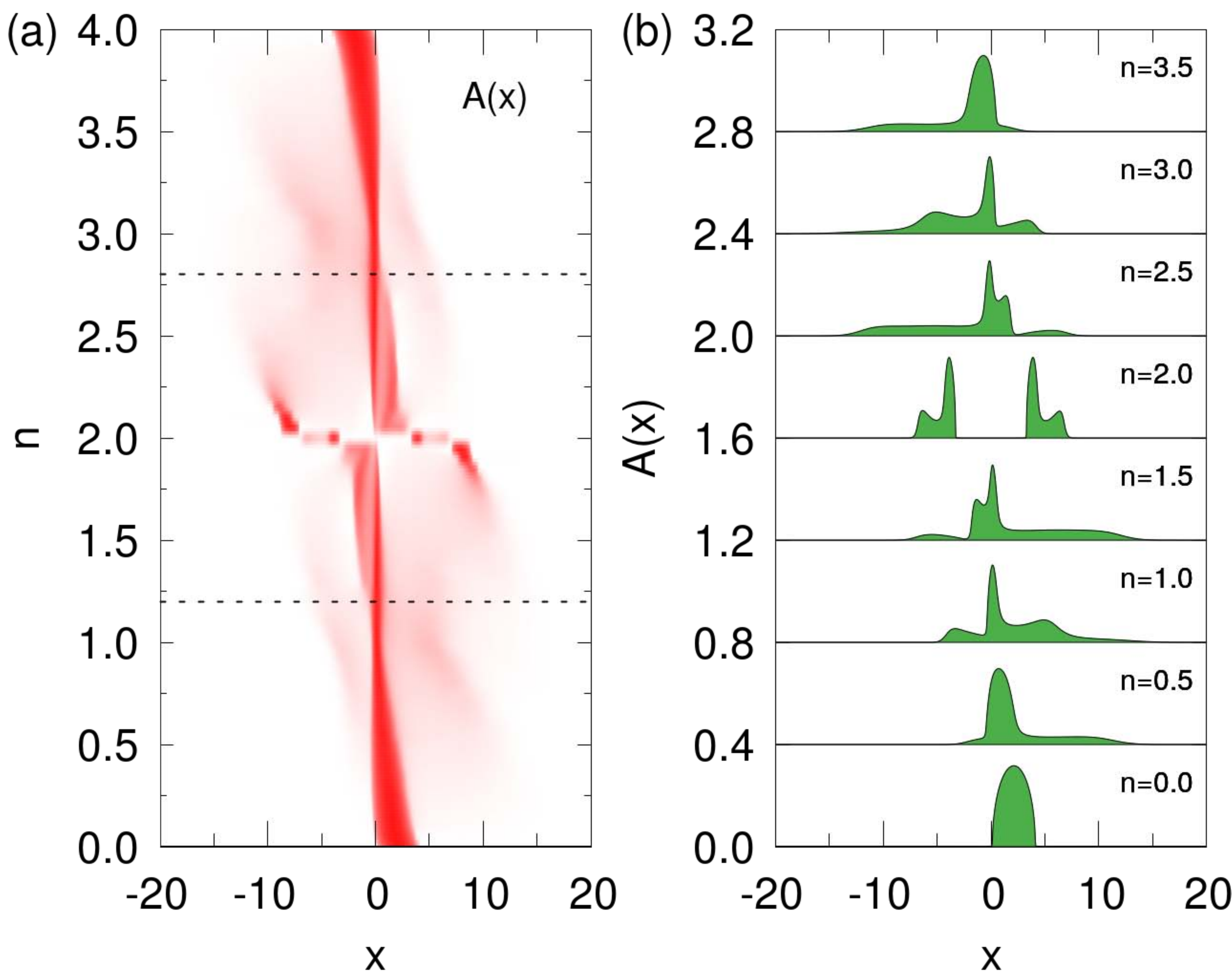}}
\vspace{-3mm}
\caption {The spectral function of the two-orbital Hubbard model as a function of the electron number $n$ per site,
obtained by applying our analytic continuation method to imaginary-frequency DMFT results.
The spectral function is plotted (a) for $0{\leq}n{\leq}4$ continuously with an intensity map and
(b) for several selected $n$.
In (b), spectral functions are offset with a step of $0.4$ for clarity.
The shoulder of the quasiparticle peak appears at about $n=1.2$ and $n=2.8$,
which are marked with dotted lines in (a).}
\vspace{-5mm}
\label{fig:6}
\end{figure}

\begin{figure}[b] 
\vspace{-3mm}
\centering
\centerline{\includegraphics[width=8.5cm]{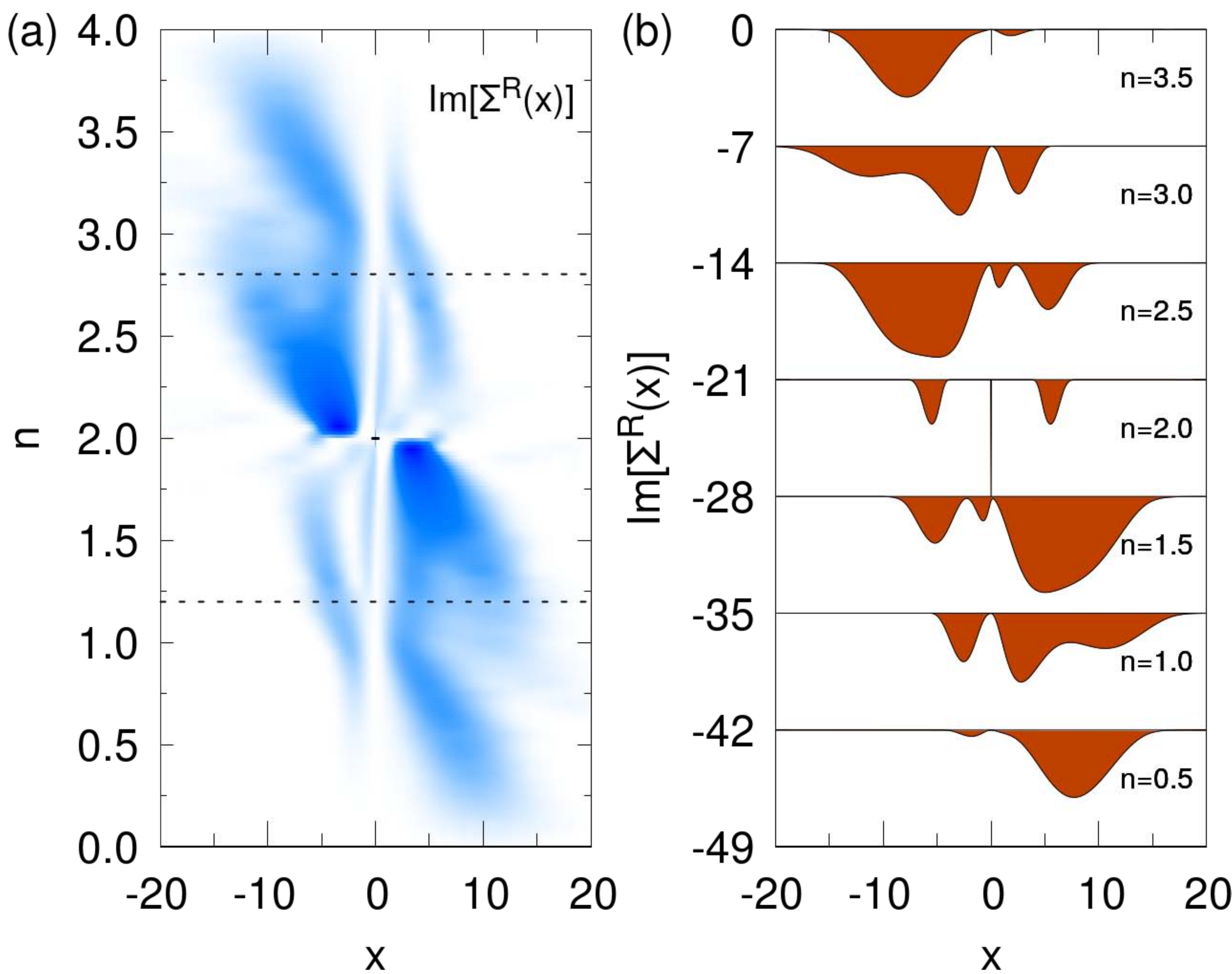}}
\vspace{-3mm}
\caption {The imaginary part of the self-energy of the two-orbital Hubbard model
as a function of the electron number $n$ per site, obtained by applying our analytic continuation method
to imaginary-frequency DMFT results.
The imaginary part of the self-energy is plotted (a) for $0{\leq}n{\leq}4$ continuously with an intensity map
and (b) for several selected $n$.
In (b), self-energies are offset with a step of $-7$ for clarity.
The self-energy diverges at the Fermi level ($x=0$) in the case of $n=2$,
which is marked with a black dot in (a).
The lower (upper) Hubbard peak splits into two peaks at about $n=1.2$ ($n=2.8$),
which is marked with a dotted line in (a).}
\label{fig:7}
\end{figure}

Figure~\ref{fig:6} shows the spectral function as a function of the electron number $n$ per site, $A(x,n)$.
The spectral function varies continuously except for the half filling ($n=2$), where the insulating phase appears.
The particle-hole symmetry, $A(x,n)=A(-x,4-n)$, is clearly observed in Fig.~\ref{fig:6},
although it is not enforced.
At $n\leq1$ or $n\geq3$, the spectral function shows a quasiparticle peak
at the Fermi level and two Hubbard bands.
As $n$ is increased from $1.2$ or decreased from $2.8$, 
a shoulder appears in the Fermi-level quasiparticle peak.

To investigate the origin of this shoulder, we plot $\operatorname{Im}[\Sigma^{R}(x,n)]$
in Fig.~\ref{fig:7}. At $n\leq1$ or $n\geq3$,
$\operatorname{Im}[\Sigma^{R}(x)]$ shows two peaks corresponding
to the lower and upper Hubbard bands, which we call the lower and upper Hubbard peaks.
As $n$ is increased from $1.2$ (decreased from $2.8$),
the lower (upper) Hubbard peak splits into two peaks,
resulting in the shoulder of the quasiparticle peak in $A(x)$.
These Hubbard-peak splittings induce non-Fermi-liquid behavior, as discussed below.

\begin{figure} 
\centering
\centerline{\includegraphics[width=8.7cm]{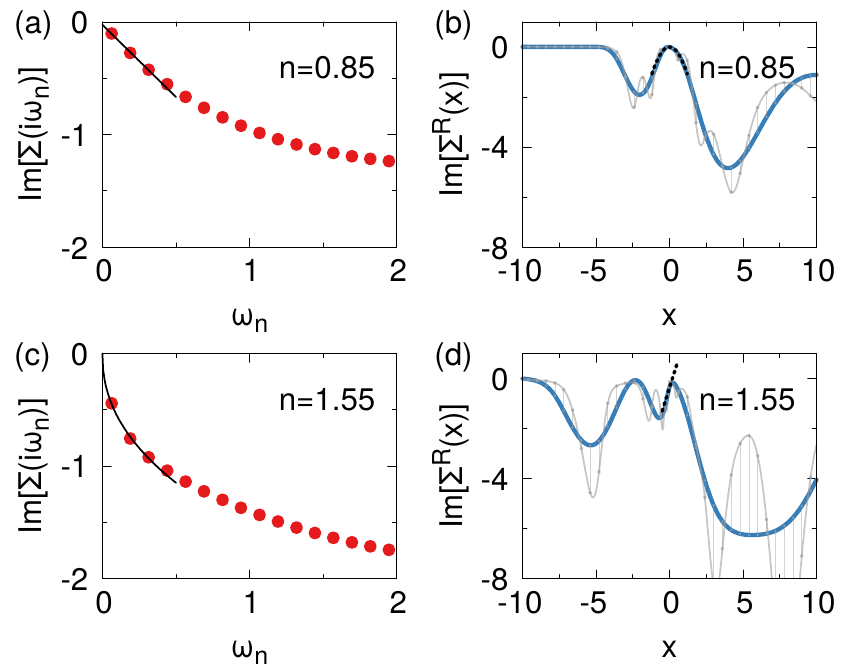}}
\vspace{-3mm}
\caption {Imaginary part of $\Sigma(i\omega_n)$, shown by red dots,
and $\Sigma^{R}(x)$, shown by blue lines, of the two-orbital Hubbard model for
(a) and (b) $n = 0.85$ and (c) and (d) $n=1.55$. In (a) and (c),
solid lines are fitted to the lowest three points of $\operatorname{Im}[\Sigma(i\omega_n)]$.
In (b) and (d), dotted lines are fitted to low-frequency part of $\operatorname{Im}[\Sigma^{R}(x)]$.
Gray lines show the deviation of $\operatorname{Im}[\Sigma^{R}(x)]$ by $\textrm{LAD}(x)$ applied to the self-energy.}
\label{fig:8}
\end{figure}

In Fig.~\ref{fig:8}, we compare the self-energies
for $n=0.85$ and $n=1.55$.
For $n = 0.85$, $\operatorname{Im}[\Sigma(i\omega_n)]$ is proportional to $\omega_n$ at small $\omega_n$, 
which indicates a Fermi-liquid behavior~\cite{Werner2008}.
In the real frequency, the Fermi-liquid behavior, $\operatorname{Im}[\Sigma^{R}(x)] = C + \alpha x^2 $
for small $x$~\cite{Imada1998}, is obtained from our analytic continuation [Fig.~\ref{fig:8}(b)].
On the other hand, for $n = 1.55$,
$\operatorname{Im}[\Sigma(i\omega_n)]$ is almost proportional 
to $\sqrt{\omega_n}$, indicating a non-Fermi-liquid behavior~\cite{Werner2008}. 
This non-Fermi-liquid behavior appears in the real frequency $x$
as linear dependance of $\operatorname{Im}[\Sigma^{R}(x)]$ on $x$ near the Fermi level ($x=0$) [Fig.~\ref{fig:8}(d)].
This linear dependence comes from splitting of Hubbard peaks in $\operatorname{Im}[\Sigma^{R}(x)]$.
Non-Fermi-liquid behavior in $\operatorname{Im}[\Sigma(i\omega_n)]$
is known to appear at the spin-freezing crossover~\cite{Werner2008,Hafermann2012,Georges2013,Hoshino2015}.
Our results show that the spin-freezing crossover
(or, equivalently, the spin-orbital separation~\cite{Stadler2015})
occurs with the Hubbard-peak splitting in $\operatorname{Im}[\Sigma^{R}(x)]$.

\begin{figure} 
\centering
\centerline{\includegraphics[width=8.5cm]{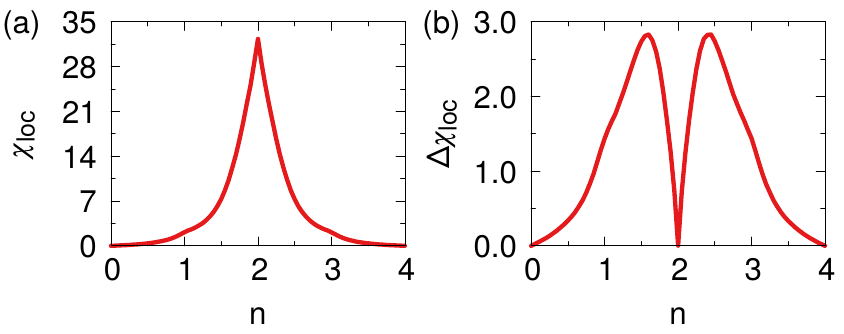}}
\vspace{-3mm}
\caption{Spin-freezing crossover of the two-orbital Hubbard model.
(a) Local magnetic susceptibility $\chi_{\textrm{loc}}$ and
(b) dynamic contribution $\Delta\chi_{\textrm{loc}}$ to the local magnetic susceptibility 
as a function of electron number $n$ per site. See the text for computational details.}
\label{fig:9}
\end{figure}

To show the spin-freezing crossover, we obtain the local magnetic susceptibility $\chi_{\textrm{loc}}$
and the dynamic contribution $\Delta\chi_{\textrm{loc}}$ to the local magnetic susceptibility.
With the operator $S_z=(1/2)\sum_{a=1}^2(n_{a\uparrow}-n_{a\downarrow})$, we define the local magnetic 
susceptibility as
\begin{equation}
    \chi_{\textrm{loc}} = \int_0^{\beta} \langle{S_z(\tau)S_z(0)}\rangle d\tau,
\end{equation}
and the dynamic contribution as
\begin{equation}
    \Delta\chi_{\textrm{loc}} = \int_0^\beta [\langle{S_z(\tau)S_z(0)}\rangle -\langle{S_z(\beta/2)S_z(0)}\rangle] d\tau.
\end{equation}
Here $\beta=1/k_BT$.
Figure~\ref{fig:9} shows $\chi_{\textrm{loc}}$ and $\Delta\chi_{\textrm{loc}}$ as functions
of the electron number $n$ per site.
As the electron number approaches the half filling ($n=2$), $\chi_{\textrm{loc}}$ increases monotonically.
On the other hand, $\Delta\chi_{\textrm{loc}}$, which represents the fluctuation of the local spin moment,
is maximal near $n = 1.6$ and $2.4$. This indicates the spin-freezing crossover~\cite{Werner2008,Hoshino2015}.

\vspace{-3mm}
\section{Summary}
\vspace{-3mm}

In summary, we developed a reliable parameter-free analytic continuation method,
tested it with exact cases, and studied the two-orbital Hubbard model as an application.
We developed a kernel grid which is suitable for the numerical analytic continuation and employed
the causal cubic spline, the second-derivative roughness penalty, and the L-curve criterion.
With these, we developed a reliable parameter-free analytic continuation method and an error estimator.
We also developed a bootstrap-averaged analytic continuation.
We demonstrated that our method reproduces the exact spectral function as statistical errors in
imaginary-frequency data decrease. 
As an application, we computed real-frequency quantities
from imaginary-frequency DMFT results of the two-orbital Hubbard model, where we
found that peaks in $\operatorname{Im}[\Sigma^{R}(x)]$
split as the electron number approaches the half filling from the quarter filling.
We verified that this peak splitting corresponds to non-Fermi-liquid behavior
considered to be the signature of the spin-freezing crossover
in previous works~\cite{Werner2008,Hafermann2012,Georges2013,Hoshino2015}.
Our analytic continuation method does not depend on any specific detail of the system
under consideration, so it can be used widely to carry out a clear real-frequency analysis
from various imaginary-time quantum many-body calculations.

\begin{acknowledgments}
\vspace{-3mm}
This work was supported by NRF of Korea (Grants No. 2020R1A2C3013673 and No. 2017R1A5A1014862) and the
KISTI supercomputing center (Project No. KSC-2021-CRE-0384).
\end{acknowledgments}

\appendix

\vspace{-3mm}
\section{\label{appendix:A}Convergence test of our kernel grid}
\vspace{-3mm}

\begin{figure} 
\centering
\centerline{\includegraphics[width=8.5cm]{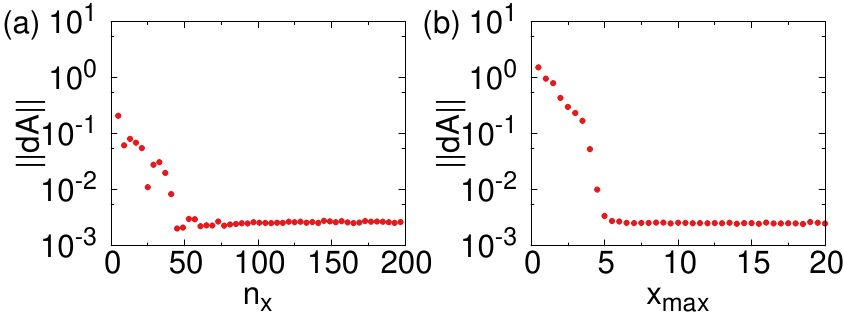}}
\vspace{-3mm}
\caption{Convergence test of the spectral function with respect to $n_x$ and $x_{\textrm{max}}$
for the case considered in Fig.~\ref{fig:2}. (a) $||dA||$ versus $n_x$. (b) $||dA||$ versus $x_\textrm{max}$. See the text for the definition of $||dA||$. In (a), $x_\textrm{max} = 10$. In (b), $n_x = 101$.}
\label{fig:10}
\end{figure}

Our analytic continuation uses the kernel grid which is a set of non-uniform $n_x$ points
in the range of $-x_{\textrm{max}}\leq x \leq x_{\textrm{max}}$, as described in the main text.
We test the convergence of the spectral function $A_{\lambda_\textrm{opt}}(x)$ calculated from our analytic continuation method
with respect to $n_x$ and $x_\textrm{max}$ by using the data and the spectral function considered in
Fig.~\ref{fig:2}. To quantify the difference between $A_{\lambda_\textrm{opt}}(x)$ and $A_{\textrm{exact}}(x)$, we define a norm
$||dA||=\{\int_{-\infty}^{\infty}(A_{\lambda_\textrm{opt}}(x)-A_{\textrm{exact}}(x))^2 dx\}^{1/2}$.
Figure~\ref{fig:10} shows $||dA||$ versus $n_x$ and $x_{\textrm{max}}$, confirming
that $A_{\lambda_\textrm{opt}}(x)$ converges to
$A_\textrm{exact}(x)$ as $n_x$ and $x_{\textrm{max}}$ increase.
At large enough $n_x$ and $x_{\textrm{max}}$,
$||dA||$ may have small nonzero values, as shown in Fig.~\ref{fig:10},
which are due to
(i) statistical errors in the imaginary-frequency data used for the 
analytic continuation and (ii) the presence of the roughness penalty $R[A]$
in Eq.~(\ref{eq:roughness}) in the regularized functional 
$Q_\lambda[A]$ of Eq.~(\ref{eq:object}).

If the Green's function $G(i\omega_n)$ is well represented with a spectral function
so that $Q_{\lambda = 0}[A]$ is minimized to a tiny value comparable to the computer precision
(as in the case of an exact Green's function without 
any statistical errors),
it is difficult to find $\lambda_\textrm{opt}$ by using the L-curve. 
In that case, it is suitable to find and use $\lambda$ at which
$Q_\lambda[A_\lambda] = 2Q_{\lambda = 0}[A_{\lambda=0}]$.

\vspace{-3mm}
\section{\label{appendix:B}Test with statistical errors proportional to
$\omega_n$}
\vspace{-3mm}

The standard deviation  of the statistical error 
in the imaginary-frequency data $G(i\omega_n)$, or $\Sigma(i\omega_n)$,
may vary with the Matsubara frequency $\omega_n$, in general.
For instance, in the two-orbital Hubbard model considered 
in Sec.~\ref{sec:2o_DMFT}, we used a quantum Monte Carlo method
to calculate the self-energy $\Sigma(i\omega_n)$, and the obtained
$\Sigma(i\omega_n)$ has larger statistical errors at larger $\omega_n$. 

\begin{figure} 
\centering
\centerline{\includegraphics[width=8.5cm]{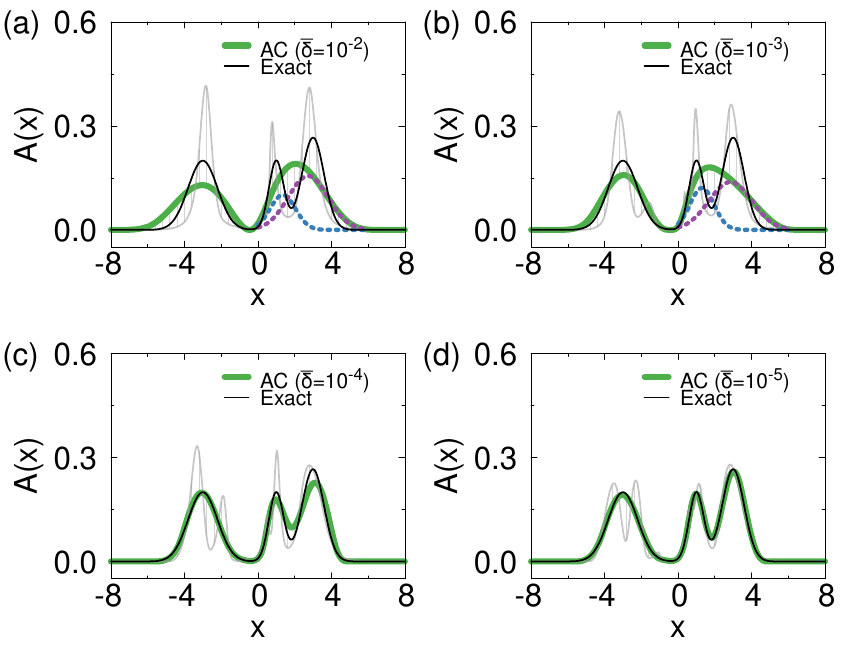}}
\vspace{-3mm}
\caption{The statistical-error dependence of the spectral function $A(x)$, shown by green lines, obtained with
our analytic continuation method.
Here $G(i\omega_n)$ have statistical errors whose standard deviation is
proportional to $\omega_n$. 
The averaged value $\bar{\delta}$ of the standard deviation is
(a) $10^{-2}$, (b) $10^{-3}$, (c) $10^{-4}$, and (d) $10^{-5}$.
Black lines show the exact spectral function, and gray lines show the deviation of $A(x)$ by $\textrm{LAD}(x)$.
In (a) and (b), Gaussian fits for $A(x)$ are shown by dotted lines.
Temperature is $0.01$. We used the first $100$ Matsubara frequencies
and a kernel grid with $x_{\textrm{max}}=10$ and $n_{x}=101$, which are large enough for converged results.}
\vspace{-3mm}
\label{fig:11}
\end{figure}

As an explicit test of the case where the statistical error varies 
with $\omega_n$,
we consider again the exact spectral function in 
Fig.~\ref{fig:4} and add Gaussian errors to the exact $G(i\omega_n)$
whose standard deviation is proportional to $\omega_n$.
This mimics the self-energy calculated by using the Dyson equation.
Let $\delta_n$ be the standard deviation of the statistical error in $G(i\omega_n)$
and $\bar{\delta}$ be the averaged value
$\bar{\delta}=\frac{1}{N_{\omega_n}}\sum_{\omega_n} \delta_{n}$.
Figure~\ref{fig:11} shows the results of our analytic continuation
with different values of $\bar\delta$,
confirming that our method works well even when the statistical error is proportional to
$\omega_n$. We also note 
$\bar{\delta}$ plays the role of $\delta$ in Fig.~\ref{fig:4}.

\begin{figure}[b] 
\centering
\centerline{\includegraphics[width=8.5cm]{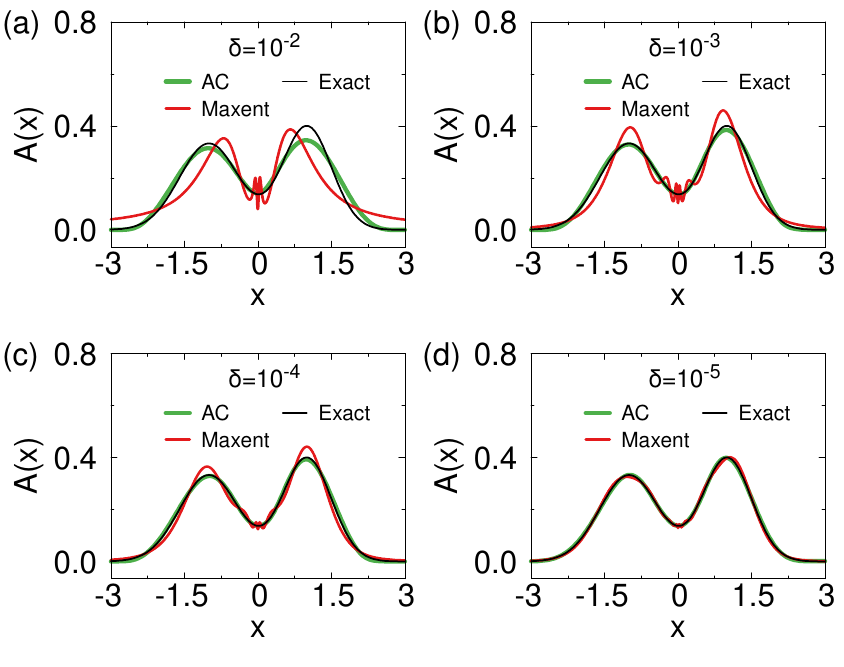}}
\vspace{-3mm}
\caption{Comparison of our method with the maximum entropy method.
Green lines are spectral functions from our analytic continuation method, and red lines 
are those from the maximum entropy method.
The standard deviation $\delta$ of the statistical error in imaginary-frequency data is
(a) $10^{-2}$, (b) $10^{-3}$, (c) $10^{-4}$, and (d) $10^{-5}$.
Black lines are exact spectral functions.
Temperature is $0.01$, and we used the first $100$ Matsubara frequencies.}
\label{fig:12}
\end{figure}

\vspace{-3mm}
\section{\label{appendix:C}Comparison with the maximum entropy method}
\vspace{-3mm}

We compare the results of our analytic continuation method with those of the maximum entropy method.
For this comparison, we implemented the maximum entropy 
method~\cite{Silver1990,Gubernatis1991,Jarrell1996}, which obtains the spectral function $A(x)$ by
minimizing $\chi^2/2-{\alpha}S[A]$.
Here $\chi^2=\sum_{i\omega_n}|G(i\omega_n)-\int\frac{A(x)}{i\omega_n-x}dx|^2$, and 
the relative entropy $S[A]$ is
\begin{equation}
  S[A] = - \int A(x) \operatorname{ln}\left(\frac{A(x)}{D(x)}\right)dx,
\end{equation}
where $D(x)$ is the default model.
The optimal value for $\alpha$ is obtained by finding the value of $\alpha$ 
that maximizes the curvature in 
the plot of $\operatorname{log}_{10}\chi^2$ versus 
$0.2 \operatorname{log}_{10}\alpha$~\cite{Bergeron2016}.
As a test example, we considered an exact spectral function $A_\textrm{exact}(x)$ which 
consists of two Gaussian peaks: one at $x = -1$ with a
standard deviation of $0.6$ and the other at $x = 1$ with a standard deviation of $0.5$.
In both our method and the maximum entropy method, we used the kernel grid 
with $x_{\textrm{max}}=10$ and $n_{x}=101$.
We generated $256$ bootstrap samples with constant Gaussian error with a standard deviation of $\delta$ which varied from $10^{-2}$ to $10^{-5}$,
and we averaged analytically continued spectral functions over bootstrap samples.
For the maximum entropy method, we used the Gaussian default model that consists of a single broad Gaussian peak at $x = 0$ with a standard deviation of $3$, and we did not apply any preblur process~\cite{Kraberger2017,Skilling1991,KaufmannPhD}.
As shown in Fig.~\ref{fig:12}, spectral functions calculated with our method and the maximum entropy 
method converge to the exact one if statistical errors are small enough [Fig.~\ref{fig:12}(d)].
If statistical errors are not small enough, the maximum entropy method produces some cusps near 
the Fermi level [Fig.~\ref{fig:12}(a)-(c)], 
as reported in the literature~\cite{Silver1990,Kraberger2017,Skilling1991,KaufmannPhD}.
While these cusps from the maximum entropy method 
become more pronounced with larger statistical errors in the imaginary-frequency data,
our method does not produce such behaviors even for large statistical-error cases.



\begin{thebibliography}{51}%
\makeatletter
\providecommand \@ifxundefined [1]{%
 \@ifx{#1\undefined}
}%
\providecommand \@ifnum [1]{%
 \ifnum #1\expandafter \@firstoftwo
 \else \expandafter \@secondoftwo
 \fi
}%
\providecommand \@ifx [1]{%
 \ifx #1\expandafter \@firstoftwo
 \else \expandafter \@secondoftwo
 \fi
}%
\providecommand \natexlab [1]{#1}%
\providecommand \enquote  [1]{``#1''}%
\providecommand \bibnamefont  [1]{#1}%
\providecommand \bibfnamefont [1]{#1}%
\providecommand \citenamefont [1]{#1}%
\providecommand \href@noop [0]{\@secondoftwo}%
\providecommand \href [0]{\begingroup \@sanitize@url \@href}%
\providecommand \@href[1]{\@@startlink{#1}\@@href}%
\providecommand \@@href[1]{\endgroup#1\@@endlink}%
\providecommand \@sanitize@url [0]{\catcode `\\12\catcode `\$12\catcode
  `\&12\catcode `\#12\catcode `\^12\catcode `\_12\catcode `\%12\relax}%
\providecommand \@@startlink[1]{}%
\providecommand \@@endlink[0]{}%
\providecommand \url  [0]{\begingroup\@sanitize@url \@url }%
\providecommand \@url [1]{\endgroup\@href {#1}{\urlprefix }}%
\providecommand \urlprefix  [0]{URL }%
\providecommand \Eprint [0]{\href }%
\providecommand \doibase [0]{https://doi.org/}%
\providecommand \selectlanguage [0]{\@gobble}%
\providecommand \bibinfo  [0]{\@secondoftwo}%
\providecommand \bibfield  [0]{\@secondoftwo}%
\providecommand \translation [1]{[#1]}%
\providecommand \BibitemOpen [0]{}%
\providecommand \bibitemStop [0]{}%
\providecommand \bibitemNoStop [0]{.\EOS\space}%
\providecommand \EOS [0]{\spacefactor3000\relax}%
\providecommand \BibitemShut  [1]{\csname bibitem#1\endcsname}%
\let\auto@bib@innerbib\@empty
\bibitem [{\citenamefont {Werner}\ \emph {et~al.}(2009)\citenamefont {Werner},
  \citenamefont {Oka},\ and\ \citenamefont {Millis}}]{Werner2009}%
  \BibitemOpen
  \bibfield  {author} {\bibinfo {author} {\bibfnamefont {P.}~\bibnamefont
  {Werner}}, \bibinfo {author} {\bibfnamefont {T.}~\bibnamefont {Oka}},\ and\
  \bibinfo {author} {\bibfnamefont {A.~J.}\ \bibnamefont {Millis}},\ }\bibfield
   {title} {\bibinfo {title} {Diagrammatic Monte Carlo simulation of
  nonequilibrium systems},\ }\href {https://doi.org/10.1103/PhysRevB.79.035320}
  {\bibfield  {journal} {\bibinfo  {journal} {Phys. Rev. B}\ }\textbf {\bibinfo
  {volume} {79}},\ \bibinfo {pages} {035320} (\bibinfo {year}
  {2009})}\BibitemShut {NoStop}%
\bibitem [{\citenamefont {Schir\'o}(2010)}]{Schiro2010}%
  \BibitemOpen
  \bibfield  {author} {\bibinfo {author} {\bibfnamefont {M.}~\bibnamefont
  {Schir\'o}},\ }\bibfield  {title} {\bibinfo {title} {Real-time dynamics in
  quantum impurity models with diagrammatic Monte Carlo},\ }\href
  {https://doi.org/10.1103/PhysRevB.81.085126} {\bibfield  {journal} {\bibinfo
  {journal} {Phys. Rev. B}\ }\textbf {\bibinfo {volume} {81}},\ \bibinfo
  {pages} {085126} (\bibinfo {year} {2010})}\BibitemShut {NoStop}%
\bibitem [{\citenamefont {Cohen}\ \emph {et~al.}(2015)\citenamefont {Cohen},
  \citenamefont {Gull}, \citenamefont {Reichman},\ and\ \citenamefont
  {Millis}}]{Cohen2015}%
  \BibitemOpen
  \bibfield  {author} {\bibinfo {author} {\bibfnamefont {G.}~\bibnamefont
  {Cohen}}, \bibinfo {author} {\bibfnamefont {E.}~\bibnamefont {Gull}},
  \bibinfo {author} {\bibfnamefont {D.~R.}\ \bibnamefont {Reichman}},\ and\
  \bibinfo {author} {\bibfnamefont {A.~J.}\ \bibnamefont {Millis}},\ }\bibfield
   {title} {\bibinfo {title} {Taming the Dynamical Sign Problem in Real-Time
  Evolution of Quantum Many-Body Problems},\ }\href
  {https://doi.org/10.1103/PhysRevLett.115.266802} {\bibfield  {journal}
  {\bibinfo  {journal} {Phys. Rev. Lett.}\ }\textbf {\bibinfo {volume} {115}},\
  \bibinfo {pages} {266802} (\bibinfo {year} {2015})}\BibitemShut {NoStop}%
\bibitem [{\citenamefont {Negele}\ and\ \citenamefont {Orland}(1988)}]{Negele}%
  \BibitemOpen
  \bibfield  {author} {\bibinfo {author} {\bibfnamefont {J.~W.}\ \bibnamefont
  {Negele}}\ and\ \bibinfo {author} {\bibfnamefont {H.}~\bibnamefont
  {Orland}},\ }\href {https://cds.cern.ch/record/729852} {\emph {\bibinfo
  {title} {{Quantum many-particle systems}}}} 
  (\bibinfo  {publisher} {Addison-Wesley},\ \bibinfo {address} {Redwood City, CA},\
  \bibinfo {year} {1988})\BibitemShut
  {NoStop}%
\bibitem [{\citenamefont {Silver}\ \emph {et~al.}(1990)\citenamefont {Silver},
  \citenamefont {Sivia},\ and\ \citenamefont {Gubernatis}}]{Silver1990}%
  \BibitemOpen
  \bibfield  {author} {\bibinfo {author} {\bibfnamefont {R.~N.}\ \bibnamefont
  {Silver}}, \bibinfo {author} {\bibfnamefont {D.~S.}\ \bibnamefont {Sivia}},\
  and\ \bibinfo {author} {\bibfnamefont {J.~E.}\ \bibnamefont {Gubernatis}},\
  }\bibfield  {title} {\bibinfo {title} {Maximum-entropy method for analytic
  continuation of quantum Monte Carlo data},\ }\href
  {https://doi.org/10.1103/PhysRevB.41.2380} {\bibfield  {journal} {\bibinfo
  {journal} {Phys. Rev. B}\ }\textbf {\bibinfo {volume} {41}},\ \bibinfo
  {pages} {2380} (\bibinfo {year} {1990})}\BibitemShut {NoStop}%
\bibitem [{\citenamefont {Vidberg}\ and\ \citenamefont
  {Serene}(1977)}]{Vidberg1977}%
  \BibitemOpen
  \bibfield  {author} {\bibinfo {author} {\bibfnamefont {H.~J.}\ \bibnamefont
  {Vidberg}}\ and\ \bibinfo {author} {\bibfnamefont {J.~W.}\ \bibnamefont
  {Serene}},\ }\bibfield  {title} {\bibinfo {title} {Solving the Eliashberg
  equations by means of N-point Pad{\'e} approximants},\ }\href
  {https://doi.org/10.1007/BF00655090} {\bibfield  {journal} {\bibinfo
  {journal} {J.Low Temp. Phys.}\ }\textbf {\bibinfo {volume}
  {29}},\ \bibinfo {pages} {179} (\bibinfo {year} {1977})}\BibitemShut
  {NoStop}%
\bibitem [{\citenamefont {Deisz}\ \emph {et~al.}(1996)\citenamefont {Deisz},
  \citenamefont {Hess},\ and\ \citenamefont {Serene}}]{Deisz1996}%
  \BibitemOpen
  \bibfield  {author} {\bibinfo {author} {\bibfnamefont {J.~J.}\ \bibnamefont
  {Deisz}}, \bibinfo {author} {\bibfnamefont {D.~W.}\ \bibnamefont {Hess}},\
  and\ \bibinfo {author} {\bibfnamefont {J.~W.}\ \bibnamefont {Serene}},\
  }\bibfield  {title} {\bibinfo {title} {Incipient Antiferromagnetism and
  Low-Energy Excitations in the Half-Filled Two-Dimensional Hubbard Model},\
  }\href {https://doi.org/10.1103/PhysRevLett.76.1312} {\bibfield  {journal}
  {\bibinfo  {journal} {Phys. Rev. Lett.}\ }\textbf {\bibinfo {volume} {76}},\
  \bibinfo {pages} {1312} (\bibinfo {year} {1996})}\BibitemShut {NoStop}%
\bibitem [{\citenamefont {Beach}\ \emph {et~al.}(2000)\citenamefont {Beach},
  \citenamefont {Gooding},\ and\ \citenamefont {Marsiglio}}]{Beach2000}%
  \BibitemOpen
  \bibfield  {author} {\bibinfo {author} {\bibfnamefont {K.~S.~D.}\
  \bibnamefont {Beach}}, \bibinfo {author} {\bibfnamefont {R.~J.}\ \bibnamefont
  {Gooding}},\ and\ \bibinfo {author} {\bibfnamefont {F.}~\bibnamefont
  {Marsiglio}},\ }\bibfield  {title} {\bibinfo {title} {Reliable Pad\'e
  analytical continuation method based on a high-accuracy symbolic computation
  algorithm},\ }\href {https://doi.org/10.1103/PhysRevB.61.5147} {\bibfield
  {journal} {\bibinfo  {journal} {Phys. Rev. B}\ }\textbf {\bibinfo {volume}
  {61}},\ \bibinfo {pages} {5147} (\bibinfo {year} {2000})}\BibitemShut
  {NoStop}%
\bibitem [{\citenamefont {Fei}\ \emph {et~al.}(2021)\citenamefont {Fei},
  \citenamefont {Yeh},\ and\ \citenamefont {Gull}}]{Fei2021}%
  \BibitemOpen
  \bibfield  {author} {\bibinfo {author} {\bibfnamefont {J.}~\bibnamefont
  {Fei}}, \bibinfo {author} {\bibfnamefont {C.-N.}\ \bibnamefont {Yeh}},\ and\
  \bibinfo {author} {\bibfnamefont {E.}~\bibnamefont {Gull}},\ }\bibfield
  {title} {\bibinfo {title} {Nevanlinna Analytical Continuation},\ }\href
  {https://doi.org/10.1103/PhysRevLett.126.056402} {\bibfield  {journal}
  {\bibinfo  {journal} {Phys. Rev. Lett.}\ }\textbf {\bibinfo {volume} {126}},\
  \bibinfo {pages} {056402} (\bibinfo {year} {2021})}\BibitemShut {NoStop}%
\bibitem [{\citenamefont {White}(1991)}]{White1991}%
  \BibitemOpen
  \bibfield  {author} {\bibinfo {author} {\bibfnamefont {S.~R.}\ \bibnamefont
  {White}},\ }\bibfield  {title} {\bibinfo {title} {The average spectrum method
  for the analytic continuation of imaginary-time data},\ }in\ \href@noop {}
  {\emph {\bibinfo {booktitle} {Computer Simulation Studies in Condensed Matter
  Physics III}}},\ \bibinfo {editor} {edited by\ \bibinfo {editor}
  {\bibfnamefont {D.~P.}\ \bibnamefont {Landau}}, \bibinfo {editor}
  {\bibfnamefont {K.~K.}\ \bibnamefont {Mon}},\ and\ \bibinfo {editor}
  {\bibfnamefont {H.-B.}\ \bibnamefont {Sch{\"u}ttler}}}\ (\bibinfo
  {publisher} {Springer},\ \bibinfo {address} {Berlin},\ \bibinfo {year} {1991}),\ pp.\ \bibinfo {pages}
  {145--153}\BibitemShut {NoStop}%
\bibitem [{\citenamefont {Sandvik}(1998)}]{Sandvik1998}%
  \BibitemOpen
  \bibfield  {author} {\bibinfo {author} {\bibfnamefont {A.~W.}\ \bibnamefont
  {Sandvik}},\ }\bibfield  {title} {\bibinfo {title} {Stochastic method for
  analytic continuation of quantum Monte Carlo data},\ }\href
  {https://doi.org/10.1103/PhysRevB.57.10287} {\bibfield  {journal} {\bibinfo
  {journal} {Phys. Rev. B}\ }\textbf {\bibinfo {volume} {57}},\ \bibinfo
  {pages} {10287} (\bibinfo {year} {1998})}\BibitemShut {NoStop}%
\bibitem [{\citenamefont {Vafayi}\ and\ \citenamefont
  {Gunnarsson}(2007)}]{Vafayi2007}%
  \BibitemOpen
  \bibfield  {author} {\bibinfo {author} {\bibfnamefont {K.}~\bibnamefont
  {Vafayi}}\ and\ \bibinfo {author} {\bibfnamefont {O.}~\bibnamefont
  {Gunnarsson}},\ }\bibfield  {title} {\bibinfo {title} {Analytical
  continuation of spectral data from imaginary time axis to real frequency axis
  using statistical sampling},\ }\href
  {https://doi.org/10.1103/PhysRevB.76.035115} {\bibfield  {journal} {\bibinfo
  {journal} {Phys. Rev. B}\ }\textbf {\bibinfo {volume} {76}},\ \bibinfo
  {pages} {035115} (\bibinfo {year} {2007})}\BibitemShut {NoStop}%
\bibitem [{\citenamefont {Ghanem}\ and\ \citenamefont
  {Koch}(2020)}]{Ghanem2020}%
  \BibitemOpen
  \bibfield  {author} {\bibinfo {author} {\bibfnamefont {K.}~\bibnamefont
  {Ghanem}}\ and\ \bibinfo {author} {\bibfnamefont {E.}~\bibnamefont {Koch}},\
  }\bibfield  {title} {\bibinfo {title} {Average spectrum method for analytic
  continuation: Efficient blocked-mode sampling and dependence on the
  discretization grid},\ }\href {https://doi.org/10.1103/PhysRevB.101.085111}
  {\bibfield  {journal} {\bibinfo  {journal} {Phys. Rev. B}\ }\textbf {\bibinfo
  {volume} {101}},\ \bibinfo {pages} {085111} (\bibinfo {year}
  {2020})}\BibitemShut {NoStop}%
\bibitem [{\citenamefont {Gubernatis}\ \emph {et~al.}(1991)\citenamefont
  {Gubernatis}, \citenamefont {Jarrell}, \citenamefont {Silver},\ and\
  \citenamefont {Sivia}}]{Gubernatis1991}%
  \BibitemOpen
  \bibfield  {author} {\bibinfo {author} {\bibfnamefont {J.~E.}\ \bibnamefont
  {Gubernatis}}, \bibinfo {author} {\bibfnamefont {M.}~\bibnamefont {Jarrell}},
  \bibinfo {author} {\bibfnamefont {R.~N.}\ \bibnamefont {Silver}},\ and\
  \bibinfo {author} {\bibfnamefont {D.~S.}\ \bibnamefont {Sivia}},\ }\bibfield
  {title} {\bibinfo {title} {Quantum Monte Carlo simulations and maximum
  entropy: Dynamics from imaginary-time data},\ }\href
  {https://doi.org/10.1103/PhysRevB.44.6011} {\bibfield  {journal} {\bibinfo
  {journal} {Phys. Rev. B}\ }\textbf {\bibinfo {volume} {44}},\ \bibinfo
  {pages} {6011} (\bibinfo {year} {1991})}\BibitemShut {NoStop}%
\bibitem [{\citenamefont {Jarrell}\ and\ \citenamefont
  {Gubernatis}(1996)}]{Jarrell1996}%
  \BibitemOpen
  \bibfield  {author} {\bibinfo {author} {\bibfnamefont {M.}~\bibnamefont
  {Jarrell}}\ and\ \bibinfo {author} {\bibfnamefont {J.}~\bibfnamefont {E.}~\bibnamefont
  {Gubernatis}},\ }\bibfield  {title} {\bibinfo {title} {Bayesian inference and
  the analytic continuation of imaginary-time quantum Monte Carlo data},\
  }\href {https://doi.org/https://doi.org/10.1016/0370-1573(95)00074-7}
  {\bibfield  {journal} {\bibinfo  {journal} {Phys. Rep.}\ }\textbf
  {\bibinfo {volume} {269}},\ \bibinfo {pages} {133} (\bibinfo {year}
  {1996})}\BibitemShut {NoStop}%
\bibitem [{\citenamefont {Gunnarsson}\ \emph {et~al.}(2010)\citenamefont
  {Gunnarsson}, \citenamefont {Haverkort},\ and\ \citenamefont
  {Sangiovanni}}]{Gunnarsson2010}%
  \BibitemOpen
  \bibfield  {author} {\bibinfo {author} {\bibfnamefont {O.}~\bibnamefont
  {Gunnarsson}}, \bibinfo {author} {\bibfnamefont {M.~W.}\ \bibnamefont
  {Haverkort}},\ and\ \bibinfo {author} {\bibfnamefont {G.}~\bibnamefont
  {Sangiovanni}},\ }\bibfield  {title} {\bibinfo {title} {Analytical
  continuation of imaginary axis data using maximum entropy},\ }\href
  {https://doi.org/10.1103/PhysRevB.81.155107} {\bibfield  {journal} {\bibinfo
  {journal} {Phys. Rev. B}\ }\textbf {\bibinfo {volume} {81}},\ \bibinfo
  {pages} {155107} (\bibinfo {year} {2010})}\BibitemShut {NoStop}%
\bibitem [{\citenamefont {Kraberger}\ \emph {et~al.}(2017)\citenamefont
  {Kraberger}, \citenamefont {Triebl}, \citenamefont {Zingl},\ and\
  \citenamefont {Aichhorn}}]{Kraberger2017}%
  \BibitemOpen
  \bibfield  {author} {\bibinfo {author} {\bibfnamefont {G.~J.}\ \bibnamefont
  {Kraberger}}, \bibinfo {author} {\bibfnamefont {R.}~\bibnamefont {Triebl}},
  \bibinfo {author} {\bibfnamefont {M.}~\bibnamefont {Zingl}},\ and\ \bibinfo
  {author} {\bibfnamefont {M.}~\bibnamefont {Aichhorn}},\ }\bibfield  {title}
  {\bibinfo {title} {Maximum entropy formalism for the analytic continuation of
  matrix-valued Green's functions},\ }\href
  {https://doi.org/10.1103/PhysRevB.96.155128} {\bibfield  {journal} {\bibinfo
  {journal} {Phys. Rev. B}\ }\textbf {\bibinfo {volume} {96}},\ \bibinfo
  {pages} {155128} (\bibinfo {year} {2017})}\BibitemShut {NoStop}%
\bibitem [{\citenamefont {Bergeron}\ and\ \citenamefont
  {Tremblay}(2016)}]{Bergeron2016}%
  \BibitemOpen
  \bibfield  {author} {\bibinfo {author} {\bibfnamefont {D.}~\bibnamefont
  {Bergeron}}\ and\ \bibinfo {author} {\bibfnamefont {A.-M.~S.}\ \bibnamefont
  {Tremblay}},\ }\bibfield  {title} {\bibinfo {title} {Algorithms for optimized
  maximum entropy and diagnostic tools for analytic continuation},\ }\href
  {https://doi.org/10.1103/PhysRevE.94.023303} {\bibfield  {journal} {\bibinfo
  {journal} {Phys. Rev. E}\ }\textbf {\bibinfo {volume} {94}},\ \bibinfo
  {pages} {023303} (\bibinfo {year} {2016})}\BibitemShut {NoStop}%
\bibitem [{\citenamefont {Wang}\ \emph {et~al.}(2009)\citenamefont {Wang},
  \citenamefont {Gull}, \citenamefont {de' Medici}, \citenamefont {Capone},\
  and\ \citenamefont {Millis}}]{Wang2009}%
  \BibitemOpen
  \bibfield  {author} {\bibinfo {author} {\bibfnamefont {X.}~\bibnamefont
  {Wang}}, \bibinfo {author} {\bibfnamefont {E.}~\bibnamefont {Gull}}, \bibinfo
  {author} {\bibfnamefont {L.}~\bibnamefont {de' Medici}}, \bibinfo {author}
  {\bibfnamefont {M.}~\bibnamefont {Capone}},\ and\ \bibinfo {author}
  {\bibfnamefont {A.~J.}\ \bibnamefont {Millis}},\ }\bibfield  {title}
  {\bibinfo {title} {Antiferromagnetism and the gap of a Mott insulator:
  Results from analytic continuation of the self-energy},\ }\href
  {https://doi.org/10.1103/PhysRevB.80.045101} {\bibfield  {journal} {\bibinfo
  {journal} {Phys. Rev. B}\ }\textbf {\bibinfo {volume} {80}},\ \bibinfo
  {pages} {045101} (\bibinfo {year} {2009})}\BibitemShut {NoStop}%
\bibitem [{\citenamefont {Kaufmann}(2021)}]{KaufmannPhD}%
  \BibitemOpen
  \bibfield  {author} {\bibinfo {author} {\bibfnamefont {J.}~\bibnamefont
  {Kaufmann}},\ }\emph {\bibinfo {title} {Local and nonlocal correlations in
  interacting electron systems}},\ \href
  {https://doi.org/10.34726/hss.2021.66083} {\bibinfo {type} {Ph.D. thesis}},\
  \bibinfo  {school} {Technische Universität Wien}, \bibinfo {year}
  {2021}\BibitemShut {NoStop}%
\bibitem [{\citenamefont {Skilling}(1991)}]{Skilling1991}%
  \BibitemOpen
  \bibfield  {author} {\bibinfo {author} {\bibfnamefont {J.}~\bibnamefont
  {Skilling}},\ }\bibfield  {title} {\bibinfo {title} {Fundamentals of maxent
  in data analysis},\ }in\ \href@noop {} {\emph {\bibinfo {booktitle} {Maximum
  Entropy in Action}}},\ \bibinfo {editor} {edited by\ \bibinfo {editor}
  {\bibnamefont {B.Buck}}\ and\ \bibinfo {editor} {\bibfnamefont
  {V.}~\bibnamefont {Macaulay}}}\ (\bibinfo  {publisher} {Clarendon},\
  \bibinfo {address} {Oxford},\ \bibinfo {year} {1991}),\ p.~\bibinfo
  {pages} {19}\BibitemShut {NoStop}%
\bibitem [{\citenamefont {Hirsch}\ and\ \citenamefont
  {Fye}(1986)}]{Hirsch1986}%
  \BibitemOpen
  \bibfield  {author} {\bibinfo {author} {\bibfnamefont {J.~E.}\ \bibnamefont
  {Hirsch}}\ and\ \bibinfo {author} {\bibfnamefont {R.~M.}\ \bibnamefont
  {Fye}},\ }\bibfield  {title} {\bibinfo {title} {Monte Carlo Method for
  Magnetic Impurities in Metals},\ }\href
  {https://doi.org/10.1103/PhysRevLett.56.2521} {\bibfield  {journal} {\bibinfo
   {journal} {Phys. Rev. Lett.}\ }\textbf {\bibinfo {volume} {56}},\ \bibinfo
  {pages} {2521} (\bibinfo {year} {1986})}\BibitemShut {NoStop}%
\bibitem [{\citenamefont {Prokof'ev}\ \emph {et~al.}(1998)\citenamefont
  {Prokof'ev}, \citenamefont {Svistunov},\ and\ \citenamefont
  {Tupitsyn}}]{Prokofev1998}%
  \BibitemOpen
  \bibfield  {author} {\bibinfo {author} {\bibfnamefont {N.~V.}\ \bibnamefont
  {Prokof'ev}}, \bibinfo {author} {\bibfnamefont {B.~V.}\ \bibnamefont
  {Svistunov}},\ and\ \bibinfo {author} {\bibfnamefont {I.~S.}\ \bibnamefont
  {Tupitsyn}},\ }\bibfield  {title} {\bibinfo {title} {Exact, complete, and
  universal continuous-time worldline Monte Carlo approach to the statistics of
  discrete quantum systems},\ }\href {https://doi.org/10.1134/1.558661}
  {\bibfield  {journal} {\bibinfo  {journal} {J. Exp. Theor. Phys.}\ }\textbf {\bibinfo {volume} {87}},\ \bibinfo {pages}
  {310} (\bibinfo {year} {1998})}\BibitemShut {NoStop}%
\bibitem [{\citenamefont {Rubtsov}\ \emph {et~al.}(2005)\citenamefont
  {Rubtsov}, \citenamefont {Savkin},\ and\ \citenamefont
  {Lichtenstein}}]{Rubtsov2005}%
  \BibitemOpen
  \bibfield  {author} {\bibinfo {author} {\bibfnamefont {A.~N.}\ \bibnamefont
  {Rubtsov}}, \bibinfo {author} {\bibfnamefont {V.~V.}\ \bibnamefont
  {Savkin}},\ and\ \bibinfo {author} {\bibfnamefont {A.~I.}\ \bibnamefont
  {Lichtenstein}},\ }\bibfield  {title} {\bibinfo {title} {Continuous-time
  quantum Monte Carlo method for fermions},\ }\href
  {https://doi.org/10.1103/PhysRevB.72.035122} {\bibfield  {journal} {\bibinfo
  {journal} {Phys. Rev. B}\ }\textbf {\bibinfo {volume} {72}},\ \bibinfo
  {pages} {035122} (\bibinfo {year} {2005})}\BibitemShut {NoStop}%
\bibitem [{\citenamefont {Werner}\ \emph {et~al.}(2006)\citenamefont {Werner},
  \citenamefont {Comanac}, \citenamefont {de' Medici}, \citenamefont {Troyer},\
  and\ \citenamefont {Millis}}]{Werner2006}%
  \BibitemOpen
  \bibfield  {author} {\bibinfo {author} {\bibfnamefont {P.}~\bibnamefont
  {Werner}}, \bibinfo {author} {\bibfnamefont {A.}~\bibnamefont {Comanac}},
  \bibinfo {author} {\bibfnamefont {L.}~\bibnamefont {de' Medici}}, \bibinfo
  {author} {\bibfnamefont {M.}~\bibnamefont {Troyer}},\ and\ \bibinfo {author}
  {\bibfnamefont {A.~J.}\ \bibnamefont {Millis}},\ }\bibfield  {title}
  {\bibinfo {title} {Continuous-Time Solver for Quantum Impurity Models},\
  }\href {https://doi.org/10.1103/PhysRevLett.97.076405} {\bibfield  {journal}
  {\bibinfo  {journal} {Phys. Rev. Lett.}\ }\textbf {\bibinfo {volume} {97}},\
  \bibinfo {pages} {076405} (\bibinfo {year} {2006})}\BibitemShut {NoStop}%
\bibitem [{\citenamefont {Gull}\ \emph {et~al.}(2008)\citenamefont {Gull},
  \citenamefont {Werner}, \citenamefont {Parcollet},\ and\ \citenamefont
  {Troyer}}]{Gull2008}%
  \BibitemOpen
  \bibfield  {author} {\bibinfo {author} {\bibfnamefont {E.}~\bibnamefont
  {Gull}}, \bibinfo {author} {\bibfnamefont {P.}~\bibnamefont {Werner}},
  \bibinfo {author} {\bibfnamefont {O.}~\bibnamefont {Parcollet}},\ and\
  \bibinfo {author} {\bibfnamefont {M.}~\bibnamefont {Troyer}},\ }\bibfield
  {title} {\bibinfo {title} {Continuous-time auxiliary-field Monte Carlo for
  quantum impurity models},\ }\href
  {https://doi.org/10.1209/0295-5075/82/57003} {\bibfield  {journal} {\bibinfo
  {journal} {{EPL}}\ }\textbf {\bibinfo {volume} {82}},\
  \bibinfo {pages} {57003} (\bibinfo {year} {2008})}\BibitemShut {NoStop}%
\bibitem [{\citenamefont {Kappl}\ \emph {et~al.}(2020)\citenamefont {Kappl},
  \citenamefont {Wallerberger}, \citenamefont {Kaufmann}, \citenamefont
  {Pickem},\ and\ \citenamefont {Held}}]{Kappl2020}%
  \BibitemOpen
  \bibfield  {author} {\bibinfo {author} {\bibfnamefont {P.}~\bibnamefont
  {Kappl}}, \bibinfo {author} {\bibfnamefont {M.}~\bibnamefont {Wallerberger}},
  \bibinfo {author} {\bibfnamefont {J.}~\bibnamefont {Kaufmann}}, \bibinfo
  {author} {\bibfnamefont {M.}~\bibnamefont {Pickem}},\ and\ \bibinfo {author}
  {\bibfnamefont {K.}~\bibnamefont {Held}},\ }\bibfield  {title} {\bibinfo
  {title} {Statistical error estimates in dynamical mean-field theory and
  extensions thereof},\ }\href {https://doi.org/10.1103/PhysRevB.102.085124}
  {\bibfield  {journal} {\bibinfo  {journal} {Phys. Rev. B}\ }\textbf {\bibinfo
  {volume} {102}},\ \bibinfo {pages} {085124} (\bibinfo {year}
  {2020})}\BibitemShut {NoStop}%
\bibitem [{\citenamefont {Efron}\ and\ \citenamefont
  {Tibshirani}(1994)}]{Efron1994}%
  \BibitemOpen
  \bibfield  {author} {\bibinfo {author} {\bibfnamefont {B.}~\bibnamefont
  {Efron}}\ and\ \bibinfo {author} {\bibfnamefont {R.}~\bibfnamefont {J.}~\bibnamefont
  {Tibshirani}},\ }\href {https://books.google.co.kr/books?id=gLlpIUxRntoC}
  {\emph {\bibinfo {title} {An Introduction to the Bootstrap}}} (\bibinfo
  {publisher} {Chapman \& Hall, New York},\ \bibinfo {year} {1993})\BibitemShut
  {NoStop}%
\bibitem [{\citenamefont {Chamandy}\ \emph {et~al.}(2012)\citenamefont
  {Chamandy}, \citenamefont {Muralidharan}, \citenamefont {Najmi},\ and\
  \citenamefont {Naidu}}]{Chamandy2012}%
  \BibitemOpen
  \bibfield  {author} {\bibinfo {author} {\bibfnamefont {N.}~\bibnamefont
  {Chamandy}}, \bibinfo {author} {\bibfnamefont {O.}~\bibnamefont
  {Muralidharan}}, \bibinfo {author} {\bibfnamefont {A.}~\bibnamefont
  {Najmi}},\ and\ \bibinfo {author} {\bibfnamefont {S.}~\bibnamefont {Naidu}},\
  }\href@noop {} {\emph {\bibinfo {title} {Estimating Uncertainty for Massive
  Data Streams}}},\ \bibinfo {type} {Technical report},\ \bibinfo  {institution}
  {Google},\ \href{https://research.google/pubs/pub43157}{https://research.google/pubs/pub43157} (\bibinfo {year} {2012})\BibitemShut {NoStop}%
\bibitem [{\citenamefont {Green}\ and\ \citenamefont
  {Silverman}(1993)}]{Green1993}%
  \BibitemOpen
  \bibfield  {author} {\bibinfo {author} {\bibfnamefont {P.}~\bibfnamefont {J.}~\bibnamefont
  {Green}}\ and\ \bibinfo {author} {\bibfnamefont {B.}~\bibfnamefont {W.}~\bibnamefont
  {Silverman}},\ }\href {https://books.google.co.kr/books?id=-AIVXozvpLUC}
  {\emph {\bibinfo {title} {Nonparametric Regression and Generalized Linear
  Models: A roughness penalty approach}}}\ (\bibinfo  {publisher} {Chapman \& Hall}, \bibinfo {adress} {London},
  \bibinfo {year} {1994})\BibitemShut {NoStop}%
\bibitem [{\citenamefont {Hansen}\ and\ \citenamefont
  {O'Leary}(1993)}]{Hansen1993}%
  \BibitemOpen
  \bibfield  {author} {\bibinfo {author} {\bibfnamefont {P.~C.}\ \bibnamefont
  {Hansen}}\ and\ \bibinfo {author} {\bibfnamefont {D.~P.}\ \bibnamefont
  {O'Leary}},\ }\bibfield  {title} {\bibinfo {title} {The use of the L-curve
  in the regularization of discrete ill-posed problems},\ }\href
  {https://doi.org/10.1137/0914086} {\bibfield  {journal} {\bibinfo  {journal}
  {SIAM J. Sci. Comput.}\ }\textbf {\bibinfo {volume} {14}},\
  \bibinfo {pages} {1487} (\bibinfo {year} {1993})}\BibitemShut {NoStop}%
\bibitem [{\citenamefont {Hansen}(2001)}]{Hansen2001}%
  \BibitemOpen
  \bibfield  {author} {\bibinfo {author} {\bibfnamefont {P.}~\bibnamefont
  {Hansen}},\ }{\selectlanguage {English}\bibinfo {title} {The L-curve and its
  use in the numerical treatment of inverse problems}},\ in\ \href@noop {}
  {{\selectlanguage {English}\emph {\bibinfo {booktitle} {Computational Inverse
  Problems in Electrocardiology}}}},\ \bibinfo {editor} {edited by\ \bibinfo
  {editor} {\bibfnamefont {P.}~\bibnamefont {Johnston}}}\ (\bibinfo
  {publisher} {WIT Press},\ \bibinfo {adress} {Southampton}\bibinfo {year} {2001}),\ pp.\ \bibinfo {pages}
  {119--142}\BibitemShut {NoStop}%
\bibitem [{\citenamefont {De~Boor}(1978)}]{Boor2}%
  \BibitemOpen
  \bibfield  {author} {\bibinfo {author} {\bibfnamefont {C.}~\bibnamefont
  {de~Boor}},\ }\href@noop {} {{\selectlanguage {English}\emph {\bibinfo
  {title} {A practical guide to splines}}}}\ (\bibinfo
  {publisher} {Springer, New York},\ \bibinfo {year} {1978})\BibitemShut {NoStop}%
\bibitem [{\citenamefont {De~Boor}(1974)}]{Boor1}%
  \BibitemOpen
  \bibfield  {author} {\bibinfo {author} {\bibfnamefont {C.}~\bibnamefont
  {de~Boor}},\ }\bibfield  {title} {\bibinfo {title} {Good approximation by
  splines with variable knots. II},\ }in\ \href@noop {} {\emph {\bibinfo
  {booktitle} {Conference on the Numerical Solution of Differential
  Equations}}},\ \bibinfo {editor} {edited by\ \bibinfo {editor} {\bibfnamefont
  {G.~A.}\ \bibnamefont {Watson}}}\ (\bibinfo  {publisher} {Springer
  },\ \bibinfo {address} {Berlin},\ \bibinfo {year}
  {1974}),\ pp.\ \bibinfo {pages} {12--20}\BibitemShut {NoStop}%
\bibitem [{\citenamefont {Ghanem}(2017)}]{GhanemPhD}%
  \BibitemOpen
  \bibfield  {author} {\bibinfo {author} {\bibfnamefont {K.}~\bibnamefont
  {Ghanem}},\ }\emph {\bibinfo {title} {{S}tochastic analytic continuation:
  {A} {B}ayesian approach}},\ \href
  {http://juser.fz-juelich.de/record/840299} {\bibinfo {type} {Ph.D. thesis}},\
  \bibinfo  {school} {RWTH Aachen University}, \bibinfo {year}
  {2017}\BibitemShut {NoStop}%
\bibitem [{\citenamefont {Lyche}\ and\ \citenamefont
  {Morken}(2008)}]{Lyche2008}%
  \BibitemOpen
  \bibfield  {author} {\bibinfo {author} {\bibfnamefont {T.}~\bibnamefont
  {Lyche}}\ and\ \bibinfo {author} {\bibfnamefont {K.}~\bibnamefont {Morken}},\
  }\href {https://cds.cern.ch/record/107441} {\emph {\bibinfo {title} {{Spline
  Methods Draft}}}}\ (\bibinfo  {publisher} {University of Oslo},\ \bibinfo
  {address} {Oslo},\ \bibinfo {year} {2008})\BibitemShut {NoStop}%
\bibitem [{\citenamefont {Nocedal}\ and\ \citenamefont
  {Wright}(2006)}]{Nocedal}%
  \BibitemOpen
  \bibfield  {author} {\bibinfo {author} {\bibfnamefont {J.}~\bibnamefont
  {Nocedal}}\ and\ \bibinfo {author} {\bibfnamefont {S.~J.}\ \bibnamefont
  {Wright}},\ }\href {https://doi.org/10.1007/978-0-387-40065-5} {\emph
  {\bibinfo {title} {{Numerical optimization, 2nd ed.}}}}\ (\bibinfo
  {publisher} {Springer},\ \bibinfo {address} {New York},\ \bibinfo {year}
  {2006})\BibitemShut {NoStop}%
\bibitem [{\citenamefont {Han}\ and\ \citenamefont {Choi}(2021)}]{mchanCO}%
  \BibitemOpen
  \bibfield  {author} {\bibinfo {author} {\bibfnamefont {M.}~\bibnamefont
  {Han}}\ and\ \bibinfo {author} {\bibfnamefont {H.~J.}\ \bibnamefont {Choi}},\
  }\bibfield  {title} {\bibinfo {title} {Causal optimization method for
  imaginary-time Green's functions in interacting electron systems},\ }\href
  {https://doi.org/10.1103/PhysRevB.104.115112} {\bibfield  {journal} {\bibinfo
   {journal} {Phys. Rev. B}\ }\textbf {\bibinfo {volume} {104}},\ \bibinfo
  {pages} {115112} (\bibinfo {year} {2021})}\BibitemShut {NoStop}%
\bibitem [{\citenamefont {Cultrera}\ and\ \citenamefont
  {Callegaro}(2020)}]{Cultrera2020}%
  \BibitemOpen
  \bibfield  {author} {\bibinfo {author} {\bibfnamefont {A.}~\bibnamefont
  {Cultrera}}\ and\ \bibinfo {author} {\bibfnamefont {L.}~\bibnamefont
  {Callegaro}},\ }\bibfield  {title} {\bibinfo {title} {A simple algorithm to
  find the L-curve corner in the regularisation of ill-posed inverse
  problems},\ }\href {https://doi.org/10.1088/2633-1357/abad0d} {\bibfield
  {journal} {\bibinfo  {journal} {{IOP} {SciNotes}}\ }\textbf {\bibinfo
  {volume} {1}},\ \bibinfo {pages} {025004} (\bibinfo {year}
  {2020})}\BibitemShut {NoStop}%
\bibitem [{\citenamefont {Hubbard}\ and\ \citenamefont
  {Flowers}(1963)}]{Hubbard1963}%
  \BibitemOpen
  \bibfield  {author} {\bibinfo {author} {\bibfnamefont {J.}~\bibnamefont
  {Hubbard}},\ }\bibfield  {title} {\bibinfo {title} {Electron correlations in
  narrow energy bands},\ }\href {https://doi.org/10.1098/rspa.1963.0204}
  {\bibfield  {journal} {\bibinfo  {journal} {Proc. R. Soc. London, Ser.\ A}\ }\textbf {\bibinfo
  {volume} {276}},\ \bibinfo {pages} {238} (\bibinfo {year}
  {1963})}\BibitemShut {NoStop}%
\bibitem [{\citenamefont {Hafermann}\ \emph {et~al.}(2012)\citenamefont
  {Hafermann}, \citenamefont {Patton},\ and\ \citenamefont
  {Werner}}]{Hafermann2012}%
  \BibitemOpen
  \bibfield  {author} {\bibinfo {author} {\bibfnamefont {H.}~\bibnamefont
  {Hafermann}}, \bibinfo {author} {\bibfnamefont {K.~R.}\ \bibnamefont
  {Patton}},\ and\ \bibinfo {author} {\bibfnamefont {P.}~\bibnamefont
  {Werner}},\ }\bibfield  {title} {\bibinfo {title} {Improved estimators for
  the self-energy and vertex function in hybridization-expansion
  continuous-time quantum Monte Carlo simulations},\ }\href
  {https://doi.org/10.1103/PhysRevB.85.205106} {\bibfield  {journal} {\bibinfo
  {journal} {Phys. Rev. B}\ }\textbf {\bibinfo {volume} {85}},\ \bibinfo
  {pages} {205106} (\bibinfo {year} {2012})}\BibitemShut {NoStop}%
\bibitem [{\citenamefont {Metzner}\ and\ \citenamefont
  {Vollhardt}(1989)}]{Metzner1989}%
  \BibitemOpen
  \bibfield  {author} {\bibinfo {author} {\bibfnamefont {W.}~\bibnamefont
  {Metzner}}\ and\ \bibinfo {author} {\bibfnamefont {D.}~\bibnamefont
  {Vollhardt}},\ }\bibfield  {title} {\bibinfo {title} {Correlated lattice
  fermions in $d=\ensuremath{\infty}$ dimensions},\ }\href
  {https://doi.org/10.1103/PhysRevLett.62.324} {\bibfield  {journal} {\bibinfo
  {journal} {Phys. Rev. Lett.}\ }\textbf {\bibinfo {volume} {62}},\ \bibinfo
  {pages} {324} (\bibinfo {year} {1989})}\BibitemShut {NoStop}%
\bibitem [{\citenamefont
  {M\"uller-Hartmann}(1989{\natexlab{a}})}]{Muller1989:a}%
  \BibitemOpen
  \bibfield  {author} {\bibinfo {author} {\bibfnamefont {E.}~\bibnamefont
  {M\"uller-Hartmann}},\ }\bibfield  {title} {\bibinfo {title} {Fermions on a
  lattice in high dimensions},\ }\href
  {https://doi.org/10.1142/S0217979289001391} {\bibfield  {journal} {\bibinfo
  {journal} {Int. J. Mod. Phys. B}\ }\textbf {\bibinfo
  {volume} {03}},\ \bibinfo {pages} {2169} (\bibinfo {year}
  {1989}{\natexlab{a}})}\BibitemShut {NoStop}%
\bibitem [{\citenamefont
  {M\"uller-Hartmann}(1989{\natexlab{b}})}]{Muller1989:b}%
  \BibitemOpen
  \bibfield  {author} {\bibinfo {author} {\bibfnamefont {E.}~\bibnamefont
  {M\"uller-Hartmann}},\ }\bibfield  {title} {\bibinfo {title} {The Hubbard
  model at high dimensions: Some exact results and weak coupling theory},\
  }\href {https://doi.org/10.1007/BF01312686} {\bibfield  {journal} {\bibinfo
  {journal} {Z. Phys. B}\ }\textbf {\bibinfo
  {volume} {76}},\ \bibinfo {pages} {211} (\bibinfo {year}
  {1989}{\natexlab{b}})}\BibitemShut {NoStop}%
\bibitem [{\citenamefont {Georges}\ and\ \citenamefont
  {Kotliar}(1992)}]{Georges1992}%
  \BibitemOpen
  \bibfield  {author} {\bibinfo {author} {\bibfnamefont {A.}~\bibnamefont
  {Georges}}\ and\ \bibinfo {author} {\bibfnamefont {G.}~\bibnamefont
  {Kotliar}},\ }\bibfield  {title} {\bibinfo {title} {Hubbard model in infinite
  dimensions},\ }\href {https://doi.org/10.1103/PhysRevB.45.6479} {\bibfield
  {journal} {\bibinfo  {journal} {Phys. Rev. B}\ }\textbf {\bibinfo {volume}
  {45}},\ \bibinfo {pages} {6479} (\bibinfo {year} {1992})}\BibitemShut
  {NoStop}%
\bibitem [{\citenamefont {Georges}\ \emph {et~al.}(1996)\citenamefont
  {Georges}, \citenamefont {Kotliar}, \citenamefont {Krauth},\ and\
  \citenamefont {Rozenberg}}]{Georges1996}%
  \BibitemOpen
  \bibfield  {author} {\bibinfo {author} {\bibfnamefont {A.}~\bibnamefont
  {Georges}}, \bibinfo {author} {\bibfnamefont {G.}~\bibnamefont {Kotliar}},
  \bibinfo {author} {\bibfnamefont {W.}~\bibnamefont {Krauth}},\ and\ \bibinfo
  {author} {\bibfnamefont {M.~J.}\ \bibnamefont {Rozenberg}},\ }\bibfield
  {title} {\bibinfo {title} {Dynamical mean-field theory of strongly correlated
  fermion systems and the limit of infinite dimensions},\ }\href
  {https://doi.org/10.1103/RevModPhys.68.13} {\bibfield  {journal} {\bibinfo
  {journal} {Rev. Mod. Phys.}\ }\textbf {\bibinfo {volume} {68}},\ \bibinfo
  {pages} {13} (\bibinfo {year} {1996})}\BibitemShut {NoStop}%
\bibitem [{\citenamefont {Werner}\ \emph {et~al.}(2008)\citenamefont {Werner},
  \citenamefont {Gull}, \citenamefont {Troyer},\ and\ \citenamefont
  {Millis}}]{Werner2008}%
  \BibitemOpen
  \bibfield  {author} {\bibinfo {author} {\bibfnamefont {P.}~\bibnamefont
  {Werner}}, \bibinfo {author} {\bibfnamefont {E.}~\bibnamefont {Gull}},
  \bibinfo {author} {\bibfnamefont {M.}~\bibnamefont {Troyer}},\ and\ \bibinfo
  {author} {\bibfnamefont {A.~J.}\ \bibnamefont {Millis}},\ }\bibfield  {title}
  {\bibinfo {title} {Spin Freezing Transition and Non-Fermi-Liquid Self-Energy
  in a Three-Orbital Model},\ }\href
  {https://doi.org/10.1103/PhysRevLett.101.166405} {\bibfield  {journal}
  {\bibinfo  {journal} {Phys. Rev. Lett.}\ }\textbf {\bibinfo {volume} {101}},\
  \bibinfo {pages} {166405} (\bibinfo {year} {2008})}\BibitemShut {NoStop}%
\bibitem [{\citenamefont {Imada}\ \emph {et~al.}(1998)\citenamefont {Imada},
  \citenamefont {Fujimori},\ and\ \citenamefont {Tokura}}]{Imada1998}%
  \BibitemOpen
  \bibfield  {author} {\bibinfo {author} {\bibfnamefont {M.}~\bibnamefont
  {Imada}}, \bibinfo {author} {\bibfnamefont {A.}~\bibnamefont {Fujimori}},\
  and\ \bibinfo {author} {\bibfnamefont {Y.}~\bibnamefont {Tokura}},\
  }\bibfield  {title} {\bibinfo {title} {Metal-insulator transitions},\ }\href
  {https://doi.org/10.1103/RevModPhys.70.1039} {\bibfield  {journal} {\bibinfo
  {journal} {Rev. Mod. Phys.}\ }\textbf {\bibinfo {volume} {70}},\ \bibinfo
  {pages} {1039} (\bibinfo {year} {1998})}\BibitemShut {NoStop}%
\bibitem [{\citenamefont {Georges}\ \emph {et~al.}(2013)\citenamefont
  {Georges}, \citenamefont {Medici},\ and\ \citenamefont
  {Mravlje}}]{Georges2013}%
  \BibitemOpen
  \bibfield  {author} {\bibinfo {author} {\bibfnamefont {A.}~\bibnamefont
  {Georges}}, \bibinfo {author} {\bibfnamefont {L.~de'}\ \bibnamefont
  {Medici}},\ and\ \bibinfo {author} {\bibfnamefont {J.}~\bibnamefont
  {Mravlje}},\ }\bibfield  {title} {\bibinfo {title} {Strong correlations from
  Hund's coupling},\ }\href
  {https://doi.org/10.1146/annurev-conmatphys-020911-125045} {\bibfield
  {journal} {\bibinfo  {journal} {Annu. Rev. Condens. Matter Phys.}\
  }\textbf {\bibinfo {volume} {4}},\ \bibinfo {pages} {137} (\bibinfo {year}
  {2013})}\BibitemShut {NoStop}%
\bibitem [{\citenamefont {Hoshino}\ and\ \citenamefont
  {Werner}(2015)}]{Hoshino2015}%
  \BibitemOpen
  \bibfield  {author} {\bibinfo {author} {\bibfnamefont {S.}~\bibnamefont
  {Hoshino}}\ and\ \bibinfo {author} {\bibfnamefont {P.}~\bibnamefont
  {Werner}},\ }\bibfield  {title} {\bibinfo {title} {Superconductivity from
  Emerging Magnetic Moments},\ }\href
  {https://doi.org/10.1103/PhysRevLett.115.247001} {\bibfield  {journal}
  {\bibinfo  {journal} {Phys. Rev. Lett.}\ }\textbf {\bibinfo {volume} {115}},\
  \bibinfo {pages} {247001} (\bibinfo {year} {2015})}\BibitemShut {NoStop}%
\bibitem [{\citenamefont {Stadler}\ \emph {et~al.}(2015)\citenamefont
  {Stadler}, \citenamefont {Yin}, \citenamefont {von Delft}, \citenamefont
  {Kotliar},\ and\ \citenamefont {Weichselbaum}}]{Stadler2015}%
  \BibitemOpen
  \bibfield  {author} {\bibinfo {author} {\bibfnamefont {K. M.}\ \bibnamefont
  {Stadler}}, \bibinfo {author} {\bibfnamefont {Z. P.}\ \bibnamefont {Yin}},
  \bibinfo {author} {\bibfnamefont {J.}~\bibnamefont {von Delft}}, \bibinfo
  {author} {\bibfnamefont {G.}~\bibnamefont {Kotliar}},\ and\ \bibinfo {author}
  {\bibfnamefont {A.}~\bibnamefont {Weichselbaum}},\ }\bibfield  {title}
  {\bibinfo {title} {Dynamical Mean-Field Theory Plus Numerical
  Renormalization-Group Study of Spin-Orbital Separation in a Three-Band Hund
  Metal},\ }\href {https://doi.org/10.1103/PhysRevLett.115.136401} {\bibfield
  {journal} {\bibinfo  {journal} {Phys. Rev. Lett.}\ }\textbf {\bibinfo
  {volume} {115}},\ \bibinfo {pages} {136401} (\bibinfo {year}
  {2015})}\BibitemShut {NoStop}%
\end{thebibliography}

%

\end{document}